\documentclass[fleqn,usenatbib]{mnras}

\usepackage{newtxtext,newtxmath}
\usepackage{multirow}

\usepackage[T1]{fontenc}

\usepackage{color}   
\usepackage{ulem}
\definecolor{dgreen}{rgb}{0,0.5,0}
\definecolor{dp}{rgb}{0.5,0,0.5}
\definecolor{magen}{rgb}{0.79,0.08,0.48}
\definecolor{darkred}{rgb}{0.65,0.06,0.37}
\usepackage{gensymb}

\DeclareRobustCommand{\VAN}[3]{#2}
\let\VANthebibliography\thebibliography
\def\thebibliography{\DeclareRobustCommand{\VAN}[3]{##3}\VANthebibliography}

\usepackage{graphicx}	
\usepackage{amsmath}	

\let\vec\mathbf

\title[Galaxy Population Properties From Quads]{Elucidating Galaxy Population Properties Using a Model-Free Analysis of Quadruply Imaged Quasar Lenses From Large Surveys} 

\author[J. H. Miller Jr and L. L. R. Williams]{
John H. Miller Jr$^{1}$\thanks{E-mail: mill9614@umn.edu (JHMJ); llrw@umn.edu (LLRW)}
and Liliya L. R. Williams$^{1}$\footnotemark[1]
\\
$^{1}$Minnesota Institute for Astrophysics, University of Minnesota, 116 Church Street, Minneapolis, MN 55455, USA 
}

\date{Accepted XXX. Received YYY; in original form ZZZ}

\pubyear{2023}

\begin{document}
\label{firstpage}
\pagerange{\pageref{firstpage}--\pageref{lastpage}}
\maketitle

\begin{abstract}
The population of strong lensing galaxies is a sub-set of intermediate-redshift massive galaxies, whose population-level properties are not yet well understood. In the near future, thousands of multiply imaged systems are expected to be discovered by wide-field surveys like Rubin Observatory's Legacy Survey of Space and Time (LSST) and Euclid. With the soon-to-be robust population of quadruply lensed quasars, or quads, in mind, we introduce a novel technique to elucidate the empirical distribution of the galaxy population properties. Our re-imagining of the prevailing strong lensing analysis does not fit mass models to individual lenses, but instead starts with parametric models of many galaxy populations, which include generally ignored mass distribution complexities and exclude external shear for now. We construct many mock galaxy populations with different properties and obtain populations of quads from each of them. The mock `observed' population of quads is then compared to those from the mocks using a model-free analysis based on a 3D sub-space of directly observable quad image properties. The distance between two quad populations in the space of image properties is measured by a metric $\eta$, and the distance between their parent galaxy populations in the space of galaxy properties is measured by $\zeta$. We find a well defined relation between $\eta$ and $\zeta$. The discovered relation between the space of image properties and the space of galaxy properties allows for the observed galaxy population properties to be estimated from the properties of their quads, which will be conducted in a future paper. 

\end{abstract}

\begin{keywords}
gravitational lensing: strong -- galaxies: fundamental parameters -- dark matter
\end{keywords}

\section{Introduction} \label{sec:intro}

The structure of the inner $5 - 15$ kpc of intermediate-redshift massive galaxies, where baryonic and dark matter co-exist, is currently not well understood. With the help of strong gravitational lensing, which is sensitive to mass fluctuations on this scale, these regions can be probed to analyze the stellar initial mass function \citep{2014posacki,2015smith,2018sonnenfeld} and the nature of dark matter \citep{2019despali,2021gilman,2023Amruth,2023vegetti}. Additionally, these galaxies serve as an evolutionary bridge between $z \sim 2$ compact galaxies and nearby early-type galaxies. 

In the near future, thousands of multiply imaged quasars and supernovae are predicted to be discovered with the Rubin Observatory's Legacy Survey of Space and Time \citep[LSST;][]{2012lsst,2010oguri,2019huber}, the Roman Space Telescope \citep{2020weiner}, the Square Kilometer Array \citep[SKA;][]{2015mckean}, and Euclid \citep{2011Laureijs}. The current population of quadruply imaged quasars, or quads, contains $N \approx 60$ quads\footnote{At present, the list of multiply imaged quasars can be found at Cameron Lemon's Gravitationally Lensed Quasar Database, https://research.ast.cam.ac.uk/lensedquasars/index.html, and C.S. Kochanek et al., CASTLES survey, https://lweb.cfa.harvard.edu/castles/} and is expected to increase by orders of magnitude. Quads are the primary lenses of choice as they provide $3\times$ more model constraints than doubles, when quasars are point sources. The monumental influx of observational data has necessitated the creation of automated search algorithms to locate strong lenses \citep{2017petrillo,2023lemon}, and methods to automate individual galaxy-lens fitting \citep{2017hezaveh,2018birrer,2019shajib,2021nightingale,2022schmidt,2023gentile}.

The standard process for strong lensing modelling is to fit a galaxy lens mass model to an individual lens by finding the model parameters that provide the best-fit to the lensing observables. Modelling individual lenses suffers from a wide range of approximate degeneracies, including the approximate mass sheet degeneracy and various shape degeneracies \citep{1985falco,2006saha}, which can lead to a biased view of lens galaxy mass distributions \citep{2021barrera,2021gomer,2023ruan, 2024etherington}. Additionally, modelling lenses independently of each other often ignores the fact that galaxy properties are governed by probability distributions, which can be correlated. For example, there exists an empirical distribution of dark matter ellipticities, which are probably correlated with the ellipticity of their stellar distribution. Analyzing the observed population of lenses simultaneously can uniquely probe these galaxy property distributions. In recent years, Bayesian hierarchical frameworks have been created to overcome this limitation of independent modelling \citep{2018sonnenfelda, 2020birrer}.

We introduce a novel technique to elucidate the empirical distribution of lens galaxy properties that uses a model-free analysis of the observed population of gravitationally lensed quads as a whole. Our analysis starts with a parameterized model of a galaxy population, from which we obtain a population of mock quads. The mock quad population is then compared to the observed quad population in a 3-dimensional space of lensing observables. The comparison is done using the distance, $\eta$ (defined in Section~\ref{sec:eta}), between the observed and mock quad populations in this 3D space. We find that this distance is correlated to the distance between their parent galaxy populations in the space of galaxy properties, $\zeta$ (defined in Section~\ref{sec:alp3}), and that this correlation extends to the origin of $\eta$ vs. $\zeta$, implying that the observed galaxy population properties can be estimated by finding the best-fit mock quad population in the 3D space of lensing observables, without mass modelling of individual lenses. Utilizing the framework laid out here, a follow-up paper will report the extracted property distributions of the observed galaxy-lens population.

An important advantage of using the observed quads as a population is that they likely can help break approximate lensing degeneracies, which arise due to insufficient data in individual lenses. With the forthcoming torrent of homogeneous gravitational lens samples, the parent galaxy-lens distributions will become better sampled with known selection biases. In the future, our analysis can be expanded to include quadruply imaged quasars created by high-resolution hydrodynamic simulations, like SEAGLE \citep{2018mukherjee}. The resulting galaxy population-level properties from our analysis can then be implemented as priors for individual lens modelling and simulations \citep{2016tacchella,2019shajib}, or hierarchical lens modelling \citep{2018sonnenfelda, 2020birrer}.

A secondary goal of our analysis is to create a parametric galaxy model that improves upon pre-existing mass models. Common mass models, like the single isothermal ellipsoid (SIE) and elliptical power-law (EPL) models, often do not accurately reconstruct lensing observables \citep{2004biggs,2006claeskens,2008jackson, 2012sluse,2019shajib,2020wagner}. To help counteract this, external shear is often included in models to account for environmental effects. However, the magnitude, direction, or both, of the external shear for specific galaxy reconstructions often differs from predictions made from the direct measurements of nearby perturbers and cosmic shear \citep{2011wong,2022cao}. In these instances, external shear is likely compensating for the lack of internal complexity in the mass model of the main lens galaxy, like boxiness, diskiness, and triaxiality \citep{1997keeton,2024etherington}.

Due to these issues, and the fact that external shear is inherently nonphysical with no associated mass, we currently exclude external shear in favor of directly modelling mass complexities. In future papers, shear will be added to account for nearby and distant massive objects (see Section \ref{sec:sum}). Our parametric galaxy model---which we use to build mock galaxy populations, and not to fit individual lenses---includes a dark matter halo, baryonic mass component, lens-plane dark matter subhalos, line-of-sight halos, and mass lopsidedness, which results from allowing an offset between centers of the dark matter halo and baryons \citep{2006gao,2019nightingale,2021barrera}. Boxiness in the mass distribution is included through the form of lensing potential we use; future work will include diskiness.

\begin{figure}
    \centering
    \includegraphics[width=0.9\columnwidth]{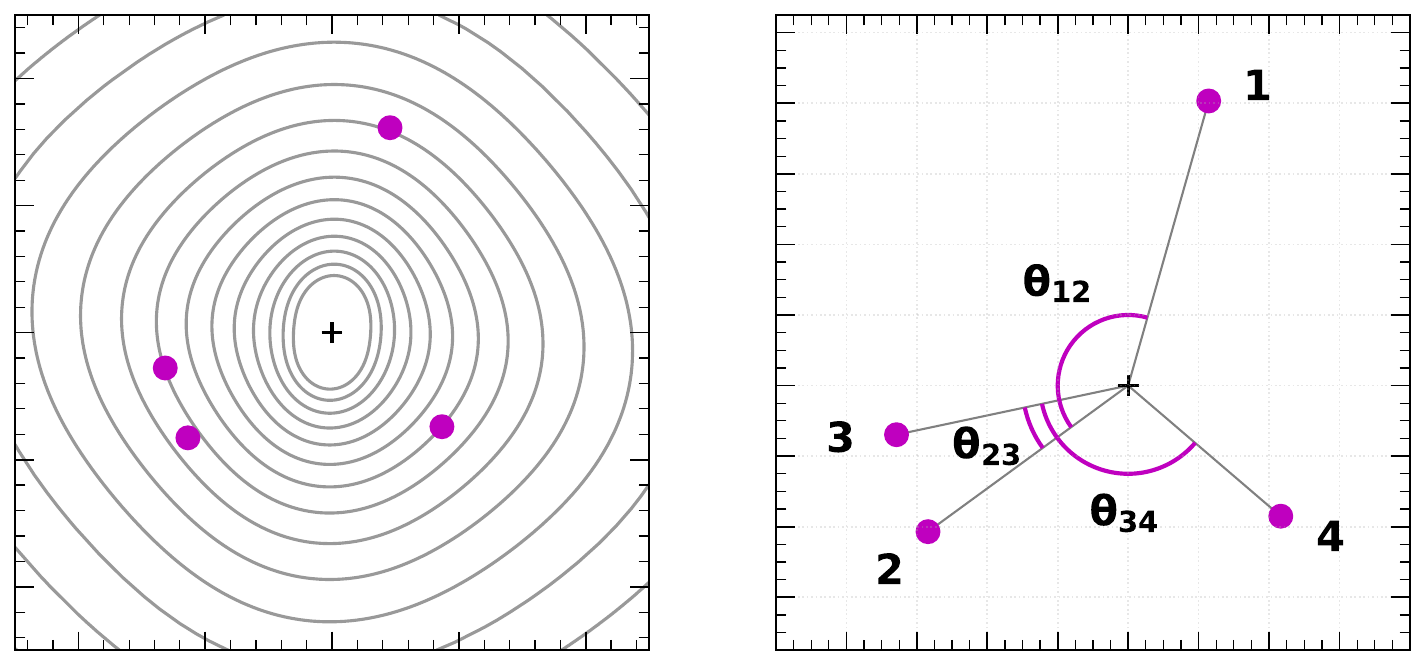}
    \caption{An example quadruply imaged quasar, as a point source. Left: The convergence field centered on the lens galaxy with the four lensed image positions. Right: The time ordering of the four images and the three defined polar angles $\theta_{12}$, $\theta_{23}$, and $\theta_{34}$. The gray lines are drawn from the center of the lens to each image.}
    \label{fig:pg1115}
\end{figure}

In Section \ref{sec:lensobs} we discuss strong gravitational lensing observables and our 3-dimensional space of observables. Section \ref{sec:mock} details the creation of our three increasingly complex parametric galaxy models that are investigated in the subsequent sections. The metric that quantifies the difference between two sets of quads, $\eta$ is introduced and assessed in Section \ref{sec:alp1} by comparing mock quad populations taken from simple, individual galaxies (what we call galaxy-galaxy comparisons). Section \ref{sec:alp2} continues galaxy-galaxy comparisons to test the efficacy of our metric, but now with more complex and realistic galaxies. In Section \ref{sec:alp3} we mimic our future comparison with the observed quad population and use our metric to distinguish between mock quad populations obtained from galaxy populations. Our analysis and results are summarized in Sections \ref{sec:sum} and \ref{sec:conc}. Although pivotal to our analysis, the reader can skip Sections \ref{sec:mods}, \ref{sec:alp1}, and \ref{sec:alp2} in order to quickly ascertain the main analysis.

\section{Lensing Observables} \label{sec:lensobs}

There are only three main categories of observables stemming from strongly lensed quasars as point sources: the image positions, image flux ratios, and time delays. The angular position $\boldsymbol{\theta}_i = (\theta_{i,x}", \theta_{i,y}")$ of the $i$th lensed image, with respect to the lensing galaxy center and with an apex at the observer, is given by the lens equation,
\begin{equation}
    {\boldsymbol\theta}_i = {\boldsymbol\alpha}_i({\boldsymbol\theta}_i) + \boldsymbol{\beta}
\end{equation}
where $\boldsymbol{\beta}$ is the angular position of the source and ${\boldsymbol\alpha}_i({\boldsymbol\theta}_i)$ the deflection angle for that image. 

The labelling of quadruply lensed images follows the time ordering of their arrival, i.e., the first arriving image = $\boldsymbol{\theta}_1$, which corresponds to the global minimum in the Fermat potential, etc. In most cases the arrival sequence of images can be deduced from the image morphology alone, where the first arriving image is usually the furthest from the center of the lens and the last arriving is the closest \citep{2003saha}. Additionally, the minima and saddle points in the Fermat potential alternate in position angle, with the first and second arriving images being minima. The two images closest to each other in proximity are the second and third arriving images. When the arrival time cannot be determined correctly by morphology alone, perturbations in the time-delay surface by galaxy substructure could be responsible \citep{2009keeton}. 

\begin{figure}
    \centering
    \includegraphics[width=0.8\columnwidth]{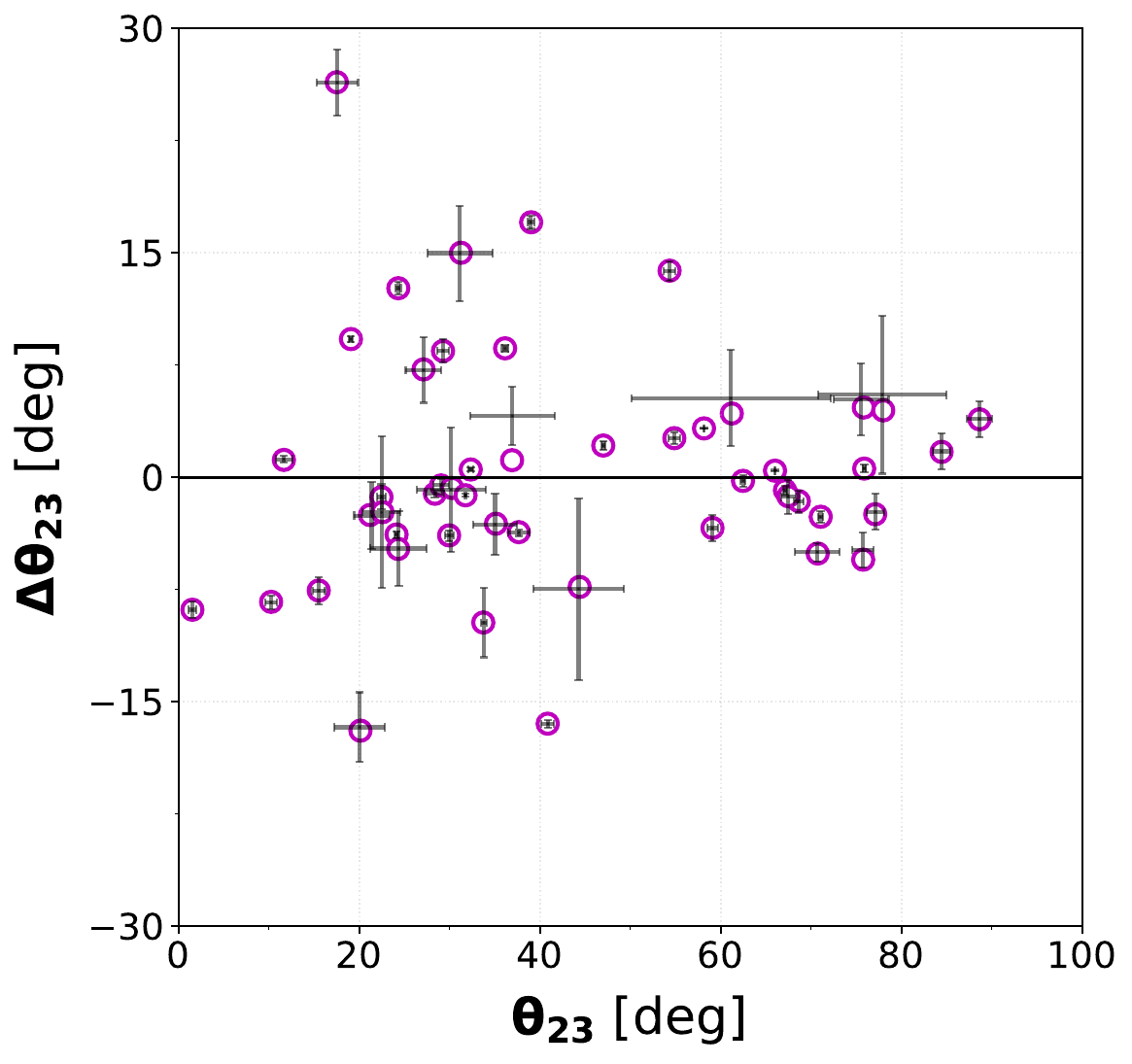}
    \caption{The 2-dimensional projection of the FSQ showing the distribution of the current sample of observed quasar quads, with the error bars determined by the astrometric uncertainty. In the cases where an error bar is not co-aligned with its point, one of the three polar angles $\theta_{12}$, $\theta_{23}$, or $\theta_{34}$ is near its upper bound and crosses it during the error determination.}
    \label{fig:obs-dt23}
\end{figure}

\begin{figure*}
    \centering
    \includegraphics[width=0.9\textwidth]{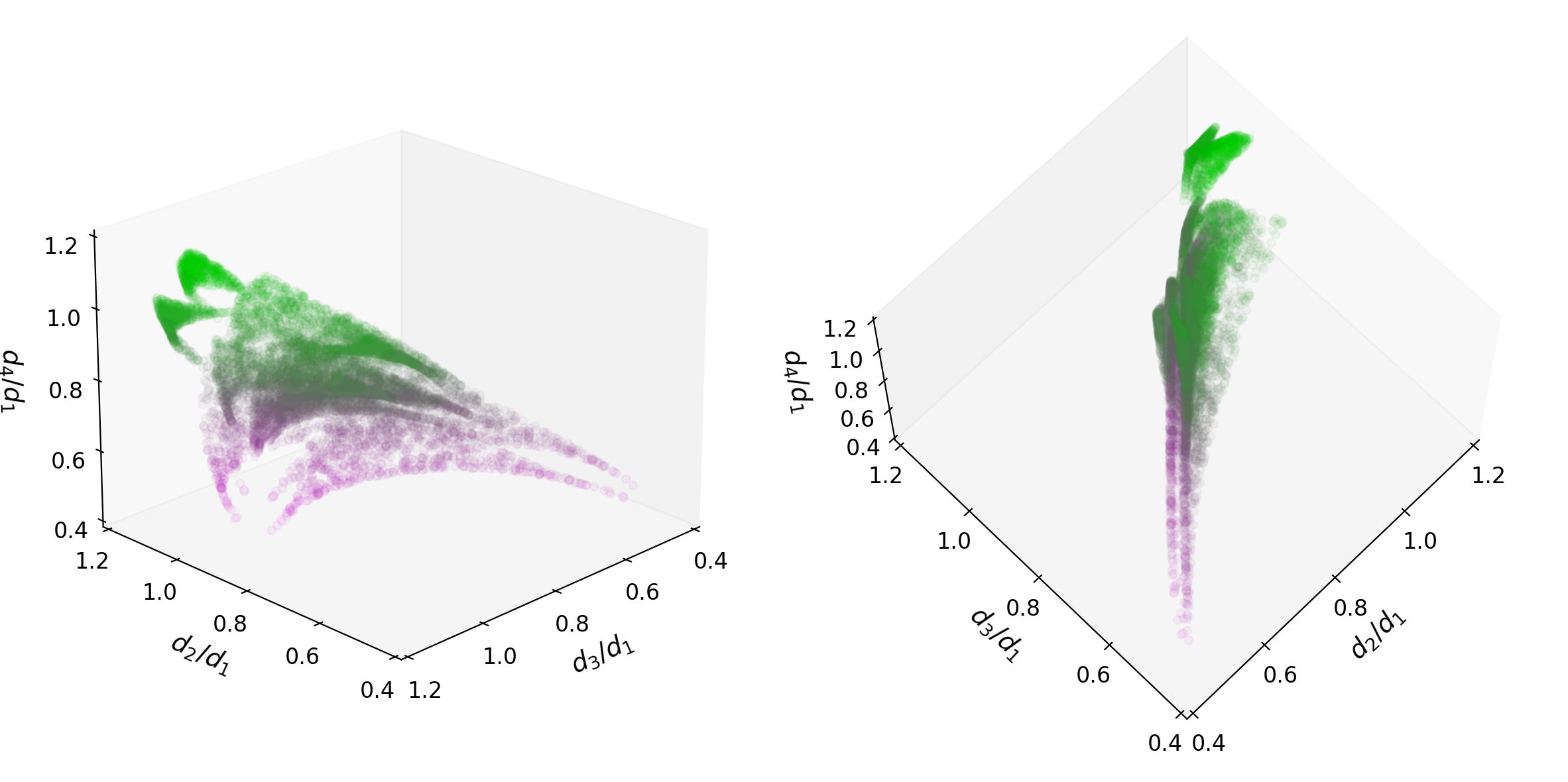}
    \caption{The 3-dimensional space of distance ratios, $d_2/d_1$, $d_3/d_1$, and $d_4/d_1$, with 10,000 quads produced by four ALPEIN-3 galaxies (discussed in Section \ref{sec:alp3}). Each point corresponds to one quad. The points are color-coded based on $d_4/d_1$: the smaller $d_4/d_1$ values are magenta and larger $d_4/d_1$ values are green. The structure shown by these quads depends on the galaxy they were lensed by and is not invariant like the FSQ. Thus, a different set of 10,000 quads will trace out different 3D shapes. }
    \label{fig:3d-dist}
\end{figure*}

Depending on the relative orientation of the images, incongruities in certain flux ratios, called `flux anomalies', can indicate the presence of substructure near the images \citep{1998mao, 2002dalal} and be used as a small-scale test of $\Lambda$CDM \citep{2007miranda,2015xu,2021gilman}. Additionally, these flux anomalies can indicate microlensing by stars or line-of-sight objects \citep{2000wozniak, 2001wambsganss, 2006dobler} or dust extinction \citep{falco1999, 2002motta, 2005mediavilla}. Or, these anomalies can be caused by the morphological properties of the lens \citep[for example, the distribution of baryons;][]{2003moller,2018hsueh,2020williams}. The two flux ratio relations are the `cusp' and `fold' relations, relating to the location of the source with respect to the caustic. When the source is inside the cusp the flux of the three clustered images should follow $f_1 + f_2 \approx f_3$, or alternatively $f_3 + f_4 \approx f_2$, where $f_i$ denotes the unsigned flux of the $i$th arriving image \citep{1998mao}. And when the source is near a fold then images 2 and 3 should follow $f_2 \approx f_3$ \citep{2005keeton}.

Initially introduced by \cite{2008williams}, the positions of the four quad images can be described by three relative polar angles with the apex at the lens center, $\theta_{12}$, $\theta_{23}$, and $\theta_{34}$ (see Fig. \ref{fig:pg1115}), where $\theta_{ij}$ describes the geometric angle between images $\boldsymbol{\theta}_i$ and $\boldsymbol{\theta}_j$ in degrees. The angle $\theta_{12}$ is between the two minima of the Fermat potential, $\theta_{34}$ between the two saddle points, and $\theta_{23}$ between the two images whose proximity to each other indicates the position of the source with respect to the caustic. We define $\theta_{12}$ and $\theta_{34}$ from ($0\degree$, $180\degree$). The angle $\theta_{23}$ is generally found from ($0\degree$, $90\degree$), but can be $> 90\degree$ on occasions due to external shear or substructure.

In a further investigation using elliptical mass distributions with arbitrary ellipticity and density slopes, these three angles were found to form a near-invariant 2-dimensional surface in 3D called ``The Fundamental Surface of Quads" \citep[FSQ;][]{2012woldesenbet}. The surface can be represented by $\theta_{23,\textup{FSQ}}$ as a function of $\theta_{12}$ and $\theta_{34}$, and thus the deviation from the FSQ is defined by $\Delta\theta_{23}$ =  $\theta_{23} - \theta_{23,\textup{FSQ}}$, where the FSQ is defined at $\Delta\theta_{23} = 0$ (see Appendix \ref{sec:fsq} for more details about the FSQ). The FSQ can be projected to 2-dimensions by plotting $\Delta\theta_{23}$ vs. $\theta_{23}$. Fig. \ref{fig:obs-dt23} shows the distribution of observed quasar quads in this projection. The magnitude of $\Delta\theta_{23}$ approximately quantifies the lens' deviation from pure ellipticity. Therefore, adding asymmetries to an elliptical mass model, like $\Lambda$CDM substructure, will result in non-zero $\Delta\theta_{23}$ for many observed quads.

In addition to the three polar angles, the image positions can be utilized to create three distance ratios. These distance ratios are found by normalizing the distance of each image from the center of the lens ($d_2$, $d_3$, and $d_4$) by the distance of the first arriving image to the center of light of the lens galaxy ($d_2/d_1$, $d_3/d_1$, and $d_4/d_1$). Typically the first arriving image is the furthest away from the lens center, and thus the distance ratios are typically $\le$ 1. 

The distance ratios of quad images produced by an elliptical mass model, when plotted in the 3D space of $d_2/d_1$, $d_3/d_1$, and $d_4/d_1$, were found to form distinct shapes (Fig. \ref{fig:3d-dist}). In contrast to the FSQ, these shapes do not seem to be invariant and instead are a more complex representation of ellipticity, density profile slope, and substructure. The quads originating from one galaxy will trace different 3D structure than quads produced from a galaxy with differing mass model. Thus, these three distance ratios provide an additional set of observables that can be utilized to probe central galaxy structure.

The power of the FSQ, i.e., just the three relative image angles, was previously demonstrated by \cite{2018gomer} to probe $\Lambda$CDM substructure in the N = 40 observed quads available at the time. The authors found that adding this substructure to an elliptical mass model produced noticeable patterns in the $\Delta\theta_{23}$ vs. $\theta_{23}$ plane, and thus deviations from the FSQ. However, the statistical effect of $\Lambda$CDM substructure could not account for observed quads' distribution of $\Delta\theta_{23}$, even when the mass of all the substructures was increased $\times 10$. That is not to say that $\Lambda$CDM substructure does not exist, or that $\Lambda$CDM is inadequate, only that it is not sufficient to explain quad image properties. Adding additional observables alongside the FSQ, like the three distance ratios, could help extract more information about galaxy structure.

In our current investigation, we utilize the 3-dimensional space of $\theta_{23}$, $\Delta\theta_{23}$, and $d_4/d_1$ lensing observables. A single quad lens corresponds to one point in this 3D space, where $\mathbf\Theta$ = ($\theta_{23}$, $\Delta\theta_{23}$, $d_4/d_1$). Our goal is to probe this 3D space to determine if the unique structure demonstrated by the 2D projection of the FSQ and one distance ratio can be utilized to constrain the inner structure of lensing galaxies (i.e., ellipticity, substructure, etc.). In other words, we aim to investigate the statistical effects of galaxy structure in this 3D space of lens observables. In turn, our ultimate goal is to constrain the lens galaxy parameters of the N $\approx$ 60 observed quads with the framework laid out here. In the future, surveys like Rubin's LSST and Euclid will deliver large uniform samples of quads that will be ideally suited for this type of analysis.

A larger set of six observables can be made with the three polar angles ($\theta_{12}$, $\theta_{23}$, and $\theta_{34}$), and three distance ratios ($d_2/d_1$, $d_3/d_1$, and $d_4/d_1$). These six describe the structural properties of the lens. A seventh parameter, the absolute scale of the quad, $d_1$, is related to the mass of the lensing galaxy and will be included in future work. However, as it will be shown, not all six or seven observables need to be used to provide compelling initial results. We limit our current investigation to three observables to determine the efficacy of our analysis. The three observables $\theta_{23}$, $\Delta\theta_{23}$, and $d_4/d_1$ were chosen as they were found to provide the most discriminating power of the six. Our analysis will be applied to the full 7D space in a future paper.

Although both time-delays and flux ratios are powerful lensing probes in their own right, we will not be using them for our analysis. Time-delays are not typically available for lensed quasars, and when they are the fractional error bars in galaxy-scale lenses are typically larger than those of the astrometry. Flux ratios, on the other hand, are sensitive to stellar microlensing, and low mass compact substructures, and probe scales smaller than our scope. Quadruply imaged quasars additionally contain a central fifth image, which can be utilized to constrain the galaxy core size \citep{2023perera}. However, since the central image is highly demagnified and not observed in galaxy scale lenses, our analysis ignores the central image. Additionally, our analysis omits extended sources, as determining what locations in the lens plane correspond to the same source can be difficult, but obvious for point sources. This correspondence is necessary to measure $\theta_{12}$, $\theta_{23}$, and $\theta_{34}$ angles.

\section{Mock Lens Galaxies \& Their Quads} \label{sec:mock}

The prevailing paradigm for strong lens modelling is to individually fit lens galaxies with their own mass model. Fitting galaxies independently of each other, as is the common practice, additionally neglects the probability distribution functions of galaxy properties, and correlations between them. Instead of mass-modelling individual galaxies, our model-free analysis is based on comparing populations of galaxies through their quad image properties in our 3D space of lensing observables, $\boldsymbol{\Theta}$ = ($\theta_{23}$, $\Delta\theta_{23}$, $d_4/d_1$). Starting with a parameterized mass model we create a population of mock galaxies, obtain a population of mock quads from them, then compare the population of observed quads to many mock quad populations in our 3D space. Our parameterized model will introduce complexities to the mass model by allowing lens plane subhalos, line-of-sight (LOS) subhalos, and lopsidedness, which is where the baryonic and dark matter components are offset and not co-aligned \citep{2006gao, 2019nightingale, 2020williams}. Perturbations from nearby and distant massive objects will be included in our future analysis (see Section \ref{sec:sum}). The form of the lensing potential we use allows for boxiness-type deviation from pure ellipticity. The magnitude of these complexities are physically motivated (discussed below) and randomized from galaxy-to-galaxy to provide for a wide range of galaxy parameter values.

\begin{table*}
    \centering
    \caption{The four different mass components, and their respective mass model (or lack there of, i.e. `-'), for the three different ALPEIN parametric galaxy models. Dark matter and baryonic matter have been shortened to DM and BM, respectively. Each mass component is represented by a letter (A-D), which is utilized in the subsequent text to help differentiate between the mass components and their respective properties. For both dark matter subhalos, each subhalo is modelled by its own Einasto profile.}
    \begin{tabular}{ccccc} \hline
        ALPEIN-\# & DM Halo (`A') & Lens Plane DM Subhalos (`B') & BM Halo (`C') & LOS DM Subhalos (`D') \\ \hline 
        ALPEIN-1 & \texttt{alphapot} & Einasto & - & - \\
        ALPEIN-2 & \texttt{alphapot} & Einasto & Einasto & Einasto \\
        ALPEIN-3 & \texttt{alphapot} & Einasto & \texttt{alphapot} & Einasto \\ \hline
    \end{tabular}
    \label{tab:models}
\end{table*}

One goal of creating our own parametric galaxy model is to explore excluding external shear in favor of directly modelling mass complexities beyond ellipticity. Commonly used mass models, like the EPL model with external shear, at times cannot reliably account for lens observable, implying that other, possibly more nuanced, lens models are sometimes needed \citep{2008jackson, 2012sluse, 2019shajib, 2020wagner, 2021barrera}. External shear is often added to account for environmental effects and line-of-sight (LOS) structures, which do affect the main lens. However, external shear can sometimes result in the biasing of galaxy parameters because some of the shear could be mimicking mass model complexity beyond ellipticity. For example, \cite{2024etherington} has shown that some portion of modelled external shear is actually mimicking boxiness. In addition to being an un-physical parameter with no associated mass, external shear was found to be unable to account for the distribution of $\Delta\theta_{23}$ in the observed population \citep{2018gomer}.

The goal of our current parameterized model is to not necessarily generate wholly realistic galaxies, per se, but to create semi-realistic galaxies that can be utilized to explore the connection between the space of lensing observables, $\Theta$, and the space of galaxy properties. 
Due to the well known degeneracy between shear and ellipticity, if our analysis was applied with the current parameterized model, which does not take into account nearby and distant massive objects, then our derived estimation for the distribution of ellipticities will be biased. In our future analysis, the parameterized galaxy model will include shear due to nearby and distant massive objects to minimize potential bias when comparing mock quad populations to the observed population of quads. 

As we show below, specific lens galaxy properties are mapped onto our 3D space of quad image observables. Thus, if a population of mock quads closely resembles the observed quad population in our 3D space, then the distribution of galaxy properties of the mock lens galaxies should also closely resemble the observed lens galaxy distribution of galaxy properties. 

\subsection{The ALPEIN Models} \label{sec:mods}

Our ALPEIN (\texttt{\textbf{alp}hapot}-\textbf{Ein}asto) parameterized galaxy model has three main iterations: ALPEIN-1, ALPEIN-2, and ALPEIN-3, with increasing complexity in the mass models. As discussed, the goal of parameterized galaxy model is to create semi-realistic galaxies, which can be utilized to explore the connection between the space of lensing observables and the space of galaxy properties. The first two model iterations, ALPEIN-1 and ALPEIN-2, corresponding to Sections \ref{sec:alp1} and \ref{sec:alp2}, respectively, are simple in their construct to motivate our later analysis. The paper's main analysis and result relies on the ALPEIN-3 mass model, which is discussed in Section \ref{sec:alp3}. The two initial models (ALPEIN-1 and 2) are included to explicate both the development of our analysis and the underlying principles responsible for our later results and conclusions. The details of the three ALPEIN models are summarized and discussed below.

All three of our galaxy models utilize a mixture of the analytical \texttt{alphapot} potential \citep{2001keeton,2011keeton} and circular Einasto profiles \citep{1965einasto}. The \texttt{alphapot} potential is a softened power law ellipsoid with five free parameters and is given by
\begin{equation} \label{eq:alphapot}
    \Phi = b \cdot \Big(s^2 + x^2 + \frac{y^2}{q^2} + K^2xy\Big)^{\alpha/2} = b \cdot \Big(s^2 + \xi^2\Big)^{\alpha/2},
\end{equation}
where $b$ is the mass normalization, $s$ the core radius, $\xi$ the elliptical radius, $\alpha$ the potential slope, and both $q^2$ and $K^2$ are parameters that define the position angle $\theta$ and axis ratio $Q$ of the potential. The position angle $\theta$ of the potential is given by \citep{2021barrera}
\begin{equation} \label{eq:pa}
    \theta = \frac{1}{2}\tan^{-1}\Bigg(\frac{q^2K^2}{q^2-1}\Bigg),
\end{equation}
and the axis ratio $Q$ of the potential is given by
\begin{equation} \label{eq:Q}
    Q^2 = \frac{\cos(2\theta)(q^2+1)+(q^2-1)}{\cos(2\theta)(q^2+1)-(q^2-1)}.
\end{equation}
Our parametric model utilizes the position angle $\theta$ and axis ratio $Q$ as input to define the \texttt{alphapot} potential. To do this, Eqs. \ref{eq:pa} and \ref{eq:Q} are rewritten to solve for $q^2$ and $K^2$, which appear in Eq. \ref{eq:alphapot}:
\begin{equation}
    q^2 = \frac{-2(Q^2+1)}{Q^2(\cos(2\theta) - 1) - (\cos(2\theta) + 1)} - 1,
\end{equation}
\begin{equation}
    K^2 = \tan(2\theta) \left(1-\frac{1}{q^2}\right).
\end{equation}
\noindent The \texttt{alphapot} potential is computationally inexpensive due to its analytical nature, a necessity when generating millions of quads. The deflection angle $\boldsymbol\alpha = \boldsymbol\nabla \Phi$ and the convergence $\kappa = 0.5 \nabla^2 \Phi$ of Eq. \ref{eq:alphapot} can be easily written and computed (see Appendix A from \citealt{2020ghosh}).

The Einasto profile is 3-dimensional circular density profile with three free parameters and is given by \citep{1965einasto, 2004navarro}
\begin{equation} \label{eq:einasto}
    \rho(r) = \rho_s ~\exp\left\{-b \left[ \left( \frac{r}{r_s} \right)^{\alpha} - 1 \right] \right\} = \rho_0  ~\exp\left\{-b \left( \frac{r}{r_s} \right)^{\alpha} \right\},
\end{equation}
\noindent where $\rho_s$ is the 3-dimensional density at the scale  radius $r_s$, $\rho_0 = \rho_s e^{b}$ is the central density, $\alpha$ the shape of the profile, and $b = 2 / \alpha$. Smaller shape parameters result in a steeper, cuspy central profile, whereas larger shape parameters result in a cored central profile. To add the Einasto profile to our galaxies, the profile is 2D projected in the lens plane \citep{2021dhar}.

All three ALPEIN models contain an elliptical dark matter halo modelled by an \texttt{alphapot} potential of mass M$_\textup{A}=10^{11}$ M$_\odot$ (labeled `A'), and many circular lens-plane dark matter subhalos modelled by Einasto profiles (labeled `B'). 
The initial ALPEIN-1 model contains only these two mass components. The later two ALPEIN models introduce a baryonic mass component (labeled `C'), modelled by an Einasto profile for ALPEIN-2 and by an \texttt{alphapot} potential for ALPEIN-3, LOS dark matter subhalos modelled by Einasto profiles (labeled `D'). Each of the four mass components are represented by a letter (A-D) to help differentiate between them. Table \ref{tab:models} summarizes these four mass components, and their respective mass models, used for the three ALPEIN models.

The ALPEIN-1 galaxy model is the simplest of the three models and contains only a dark matter halo and three dark matter subhalos, with a subhalo mass fraction upper limit of 10\%. The goal of this initial model is not to generate physical galaxies, per se, but instead to provide a straightforward preliminary analysis of our 3D space. Four ALPEIN-1 sub-models are created to test connections between the subhalo modelling parameters and our observables. These sub-models and their results are discussed in Section \ref{sec:alp1}. 

In an effort to generate more physically motivated galaxies, the succeeding ALPEIN-2 model introduces an Einasto baryonic component (`C') and Einasto LOS subhalos (`D') to the existing ALPEIN-1 model. Instead of only allowing three lens-plane subhalos, both lens-plane and LOS subhalos are numerous in number and limited by the total mass allocated to them. The combined upper mass limits of the lens-plane and LOS subhalos, M$_\textup{BD}^\textup{lim}$, is chosen randomly between $0 - 10\%$ of the main dark matter component mass, and the ratio M$_\textup{B}^\textup{lim}$ $/$ M$_\textup{BD}^\textup{lim}$ is uniformly distributed between $3/12 - 5/12$, with a mean of $\approx 1/3$. There is twice as much mass along the line-of-sight as compared to the lens plane. This proportion already incorporates $\kappa$ re-scaling due to $z_\textup{subhalo}$ \citep{2022he}. The subhalo masses are given by the \cite{2008springel} subhalo mass function, with the smallest subhalo mass, m$_\textup{min}$, being $10^8$ M$_\odot$. The lens-plane subhalos are distributed randomly in polar coordinates, and randomly in radius, whereas LOS subhalos are distributed randomly in Cartesian coordinates. Both components have a maximum distance of 4 Einstein radii (20 kpc) away from the center of the baryonic component, and the lens-plane subhalos have a minimum distance of 0.1 $\theta_E$ (0.5 kpc).

\begin{table*}
    \centering
    \caption{The mass model parameters for the four different ALPEIN-1 models. The first three parameters, $s, Q,$ and $\alpha_\textup{A}$ are for the \texttt{alphapot} (Eq. \ref{eq:alphapot}) dark matter halo. The last five parameters, r$_s$, $\rho_0$, $\alpha_\textup{B}$, N, and Positions, are for the Einasto (Eq. \ref{eq:einasto}) dark matter subhalos. The dark matter halo potential slope, $\alpha_\textup{A}$ = 1, is isothermal. The subhalos' shape parameters are chosen to be large (i.e., $\alpha_\textup{B} = 3$) for our initial test.}
    \begin{tabular}{ccccccccccc} \hline
        & \multicolumn{3}{c}{DM Halo (`A')} & \multicolumn{5}{c}{DM Subhalos (`B')} \\ \hline
        \multirow{2}{*}{ALPEIN-\#} & $s$ & $Q$ & $\alpha_\textup{A}$ & $r_s$ & $\rho$ & $\alpha_\textup{B}$ & N & Positions \\
        & [kpc] & & & [kpc] & [g cm$^{-3}$] & & & \\ \hline
        ALPEIN-1-1 & 0.275 & 1.2 & 1.0 & 0.125 $-$ 1 & $10^{-2.3}$ & 3 & 3 & Constant \\
        ALPEIN-1-2 & 0.275 & 1.2 & 1.0 & 0.625 &  $10^{-3} - 10^{-1.5}$ & 3 & 3 & Constant \\
        ALPEIN-1-3 & 0.275 & 1.2 & 1.0 & 0.125 $-$ 1.75 & $10^{-3} - 10^{-1.5}$ & 3 & 3 & Constant \\
        ALPEIN-1-4 & 0.275 & 1.2 & 1.0 & 0.125  $-$ 1.75 & $10^{-3} - 10^{-1.5}$ & 3 & 3 & Random \\ \hline
    \end{tabular}
    \label{tab:alpein1}
\end{table*}

\begin{figure}
    \centering
    \includegraphics[width=0.9\columnwidth]{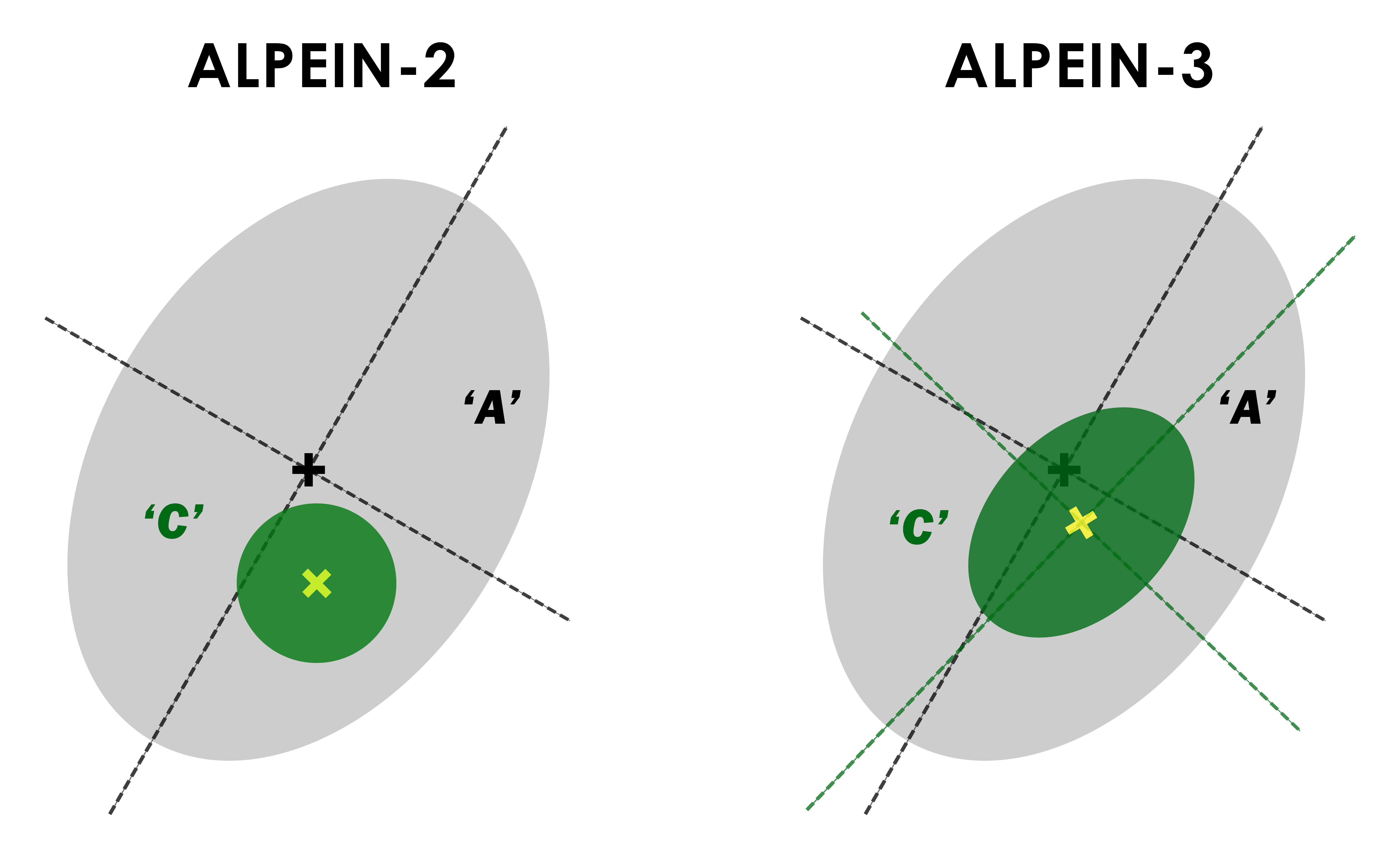}
    \caption{Diagrams demonstrating how lopsidedness is implemented in ALPEIN-2 and ALPEIN-3, respectively. The gray oval depicts the main dark matter halo, labelled `A' with center `+', and the green circle (ALPEIN-2; Left) and oval (ALPEIN-3; Right) depicts the baryonic component, labelled `C' with center `x'. For ALPEIN-2, only the position of the baryonic component determines the lopsidedness. In ALPEIN-3, the position, position angle, and ellipticity are allowed to vary for more complex lopsidedness. Offsets are exaggerated for clarity.}
    \label{fig:alp3-off}
\end{figure}

\begin{figure*}
    \centering
    \includegraphics[width=0.91\textwidth]{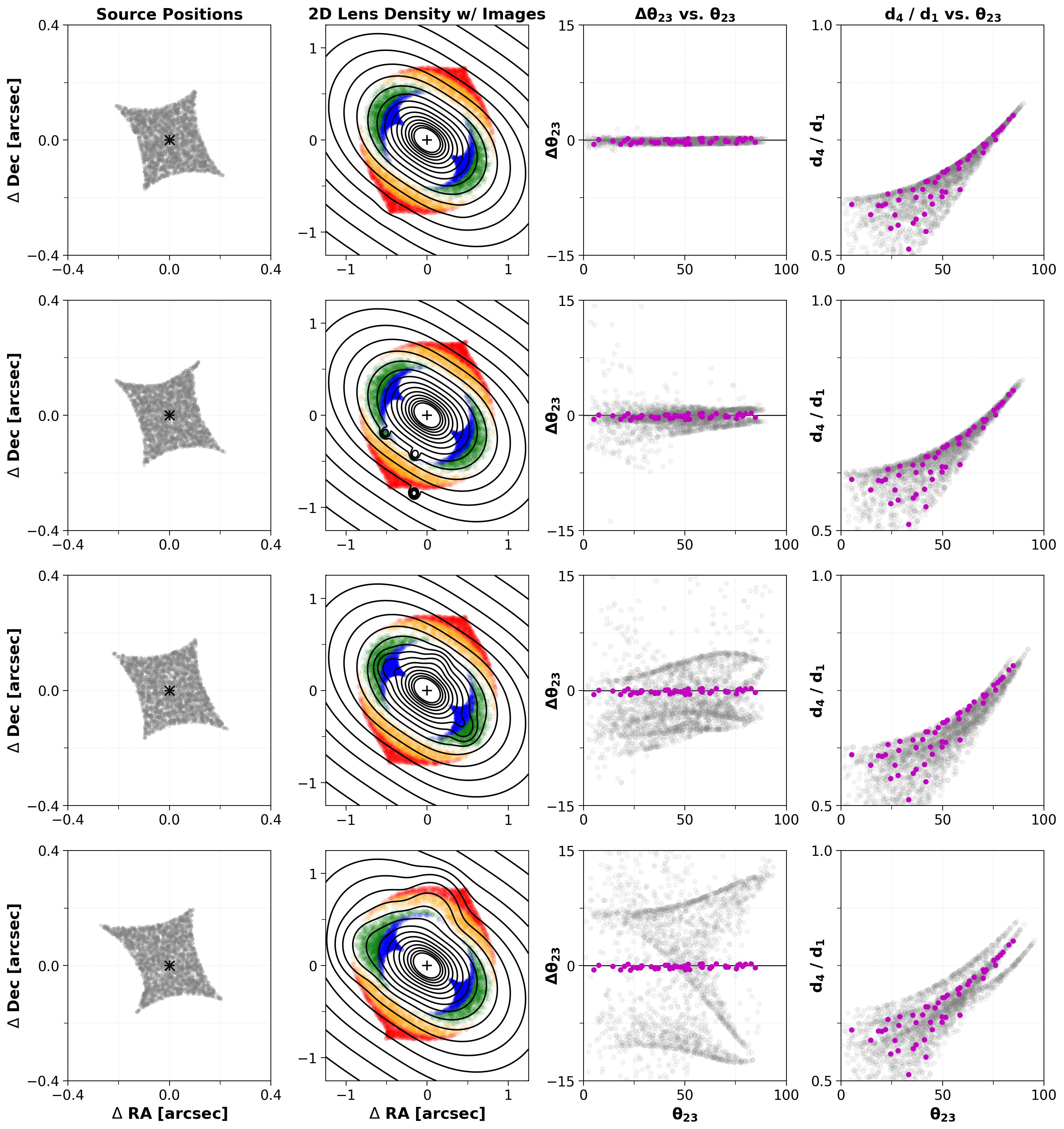}
    \caption{Examples of 4 ALPEIN-1-4 galaxy lenses (Rows 1-4). The main dark matter halo is constant for all galaxies. Each of the three subhalos in the same galaxy have the same total mass, but this mass differs from galaxy-to-galaxy. Their positions also differ. Column 1 shows randomly sampled source positions, where `+' is the center of the dark matter halo.  Column 2 shows the corresponding 2D lens density with the first arriving images in red, second in yellow, third in green, and fourth in blue. Column 3 shows the comparison between the N=50 observed quads (magenta) taken from the first galaxy (Row 1) and M=2450 mock quads (gray) in the 2D projection of the FSQ, $\Delta\theta_{23}$ vs. $\theta_{23}$. Column 4 shows the comparison between the same observed and mock quads, but this time in $d_4/d_1$ vs. $\theta_{23}$ plane. The galaxies are arranged in order of increasing subhalo mass fraction from top to bottom.}
    \label{fig:alp1}
\end{figure*}

The LOS subhalos are modelled in the galaxy lens-plane. Perturbations along the LOS can occur anywhere between the lens and the observer, or between the lens and the source. Correctly modelling these perturbations depends on the recursive multiplane lens modelling formalism (see Fig. 6 of \citealt{2019gilman}). To lowest order, one can model the LOS subhalos in the lens plane (with re-scaling), and for our purposes that is sufficient. 

Elliptical galaxies can exhibit azimuthal asymmetry in their stellar mass distribution in the form of lopsidedness, boxiness, diskiness, etc. These could be the result of asymmetric mass accretion resulting in the galaxy not being dynamically relaxed. Lopsidedness near the Einstein radius can be modelled by allowing the centers of the two main mass components, A and C, to be misaligned \citep[e.g.,][see Fig. \ref{fig:alp3-off}]{2021barrera}. 

Elliptical galaxies from SLACS \citep{2019nightingale} and from \cite{2019shajib} were found to have offsets between the centers of observed light and modeled mass distribution of up to $\approx 0.3 - 0.5$ kpc. To allow for a wide range of galaxy characteristics, we adopt a maximum offset of 0.5 kpc for the two main mass components. We note that even the largest offsets produce galaxies that have a single combined mass center (see the second column in Fig.~\ref{fig:alp2}.)

To be consistent with observations the center of the lens, and thus the apex of the three relative angles, is set to be the center of the baryonic component, C. The mass of this baryonic component is 10\% of the main dark matter halo and is modelled by a circular Einasto profile. The lensing potential axial-ratios of the main dark matter halo, $Q_\textup{A}$, are drawn randomly from the convergence (surface mass density) axial-ratios, $\epsilon$, taken from \citealt[][Table 5]{2008bolton}, with the relation $Q_\textup{A} \approx \epsilon^{1/2.75}$ \citep{2023lasko}. 

The distribution of baryonic matter in elliptical galaxies often has non-zero ellipticity and cannot be represented by the projected Einasto profile, of which no simple elliptical version exists. Thus, ALPEIN-3 improves on ALPEIN-2 by modelling the baryonic component with an elliptical \texttt{alphapot} potential. The axial ratio of the baryonic component, $Q_\textup{C}$, is allowed to vary $\pm$ 0.2 from the main dark matter halo, $Q_\textup{A}$, and its position angle $\theta_\textup{C}$ $\pm 30\degree$ from $\theta_\textup{A}$ (see Fig. \ref{fig:alp3-off}). The baryonic component mass is increased to $3 \times 10^{10}$ M$_{\odot}$, and its normalization factor b$_\textup{C}$ is chosen such that the fraction of mass contained inside the Einstein radius $\theta_E$ is M$_\textup{C}$ $/$ M$_\textup{A}$ $\approx 30\%$ \citep{2005ferreras,2015vincente}. As for the subhalos, their generation remains the same from ALPEIN-2 to ALPEIN-3. 

Mock galaxies are generated from the parameterized mass model by randomizing select galaxy parameters, which depends on their ALPEIN model. Once the galaxy has been created, a population of quads is obtained from it. Quads are generated by randomly picking source positions until five image configurations are found and are ordered based on the Fermat potential for consistency. Throughout our parameterized modelling we use a lens redshift z$_l = 0.4$ and source redshift z$_s$ = 3.0.

In Section \ref{sec:alp1} we use the basic ALPEIN-1 mass model to create a metric, $\eta$, in our 3-dimensional space that compares quad populations lensed by individual galaxies (i.e., galaxy-galaxy comparisons). We test the effectiveness our $\eta$ metric in Section \ref{sec:alp2} with quad populations lensed by complex galaxies generated from the ALPEIN-2 mass model. Finally, with ALPEIN-3 in Section \ref{sec:alp3} we test population-population comparisons, which is where a population of quads is generated by a population of galaxies, and serves as our main result.

\section{ALPEIN-1: Initial Galaxy-Galaxy Comparisons} \label{sec:alp1}

To start the analysis of our 3D space of lensing observables, we create the very simple mass model ALPEIN-1. The model has two components: a dark matter halo (`A') modelled by an elliptical \texttt{alphapot} potential and three dark matter subhalos (`B') modelled by circular Einasto profiles. As mentioned before, the angle $\Delta\theta_{23}$ approximately quantifies the deviation from ellipticity for a given mass distribution. Adding asymmetry in a mass distribution in the form of substructure will result in non-zero values of $\Delta\theta_{23}$, which increase in magnitude with increasing subhalo mass \citep{2018gomer}. The simplicity of our ALPEIN-1 mass model provides a straightforward environment to analyze this connection between subhalo parameters and our 3D space of observables, $\theta_{23}$, $\Delta\theta_{23}$, and $d_4/d_1$, and to develop a metric to quantify this connection.

To investigate this connection we create four ALPEIN-1 sub-models, ALPEIN-1-1, ALPEIN-1-2, ALPEIN-1-3, and ALPEIN-1-4, which all vary different subhalo parameters. The dark matter halo (`A') parameters are the same for each sub-model and are kept constant for all mock galaxies. All three dark matter subhalos (`B') in a given galaxy have the same subhalo parameter values (r$_s$, $\rho_0$, and $\alpha$), but one of these properties is varied galaxy-to-galaxy depending on the sub-model. ALPEIN-1-1 randomizes the scale radius r$_s$ between mock galaxies, ALPEIN-1-2 randomizes the central density $\rho_0$ between mock galaxies, and both ALPEIN-1-3 and ALPEIN-1-4 randomize both r$_s$ and $\rho_0$ between mock galaxies. ALPEIN-1-4 randomizes the three subhalo positions inside a small annulus containing the Einstein radius. Whereas each galaxy in the first three ALPEIN-1 sub-models have constant subhalo locations. The sub-model properties are summarized in Table \ref{tab:alpein1}. 

We only keep galaxies with subhalo mass fractions $0.0001 < \textup{M}_\textup{B} / \textup{M}_\textup{A} < 0.1$ within 6.25 kpc from the center of the lens \citep{2008springel}. The ranges in Table \ref{tab:alpein1} are initial ranges. If, for example, large r$_s$ and $\rho_0$ values result in a mass fraction exceeding our mass fraction limit, then the galaxy is not kept.

A population of 200 randomized mock galaxies were generated from each of the ALPEIN-1 sub-models. From each of these mock galaxies, a population of 2500 quads were obtained, resulting in half a million quads per mass model. Figure \ref{fig:alp1}, column 2 shows four example galaxies from ALPEIN-1-4 and their quads, which are sorted from lowest mass fraction (top) to highest mass fraction (bottom).

As a precursor to our full 3-dimensional comparisons (Section \ref{sec:eta}), we show how the deviation from the FSQ, or $\Delta\theta_{23}$, is affected by subhalo mass (Column 3 of Fig.~\ref{fig:alp1}). By increasing our subhalo Einasto parameters, r$_s$ and $\rho_0$, the subhalo mass fraction increases, which results in larger magnitudes of $\Delta\theta_{23}$ (gray points in Column 3 of Fig. \ref{fig:alp1}). Furthermore in Column 3, which is the 2D projection of the FSQ, we find that subhalos create unique lines, or filaments, curving through this 2-dimensional space. Due to having only three subhalos and a high mass fraction limit, the most massive subhalos are unrealistic, but are kept to help analyze the impact of subhalo mass. The following two mass models, ALPEIN-2 and ALPEIN-3 (Sections \ref{sec:alp2} and \ref{sec:alp3}), model subhalos more realistically, and thus the impact of them on $\Delta\theta_{23}$ will decrease. The observables $\theta_{23}$ and $d_4/d_1$ did not show any direct correlations to subhalo parameters, but did result in 2-dimensional structure between the two observables (gray points in Column 4 in Fig.~\ref{fig:alp1}). To quantify the similarity, or dissimilarity, of these structures requires the creation of a metric.

\subsection{The Comparison Metric, \texorpdfstring{$\eta$}{eta}} \label{sec:eta}

We create the metric $\eta$, which finds the average 3-dimensional distance between two sets of quad populations of size $N$ and $M$, respectively, in our 3D space of lensing observables. To mirror our future analysis with the true observed quad population, we choose one population to be designated as the `observed' mock population, and to contain $N = 50$ quads (the then current number of quads, updated to $N = 60$ for Section \ref{sec:alp3}), and the other as the mock population containing $M$ number of quads. To simplify verbiage, in the rest of the paper we call the `observed' mock population as simply the observed population, keeping in mind that these are not real observed lenses.

\begin{figure}
    \centering
    \includegraphics[width=1\columnwidth]{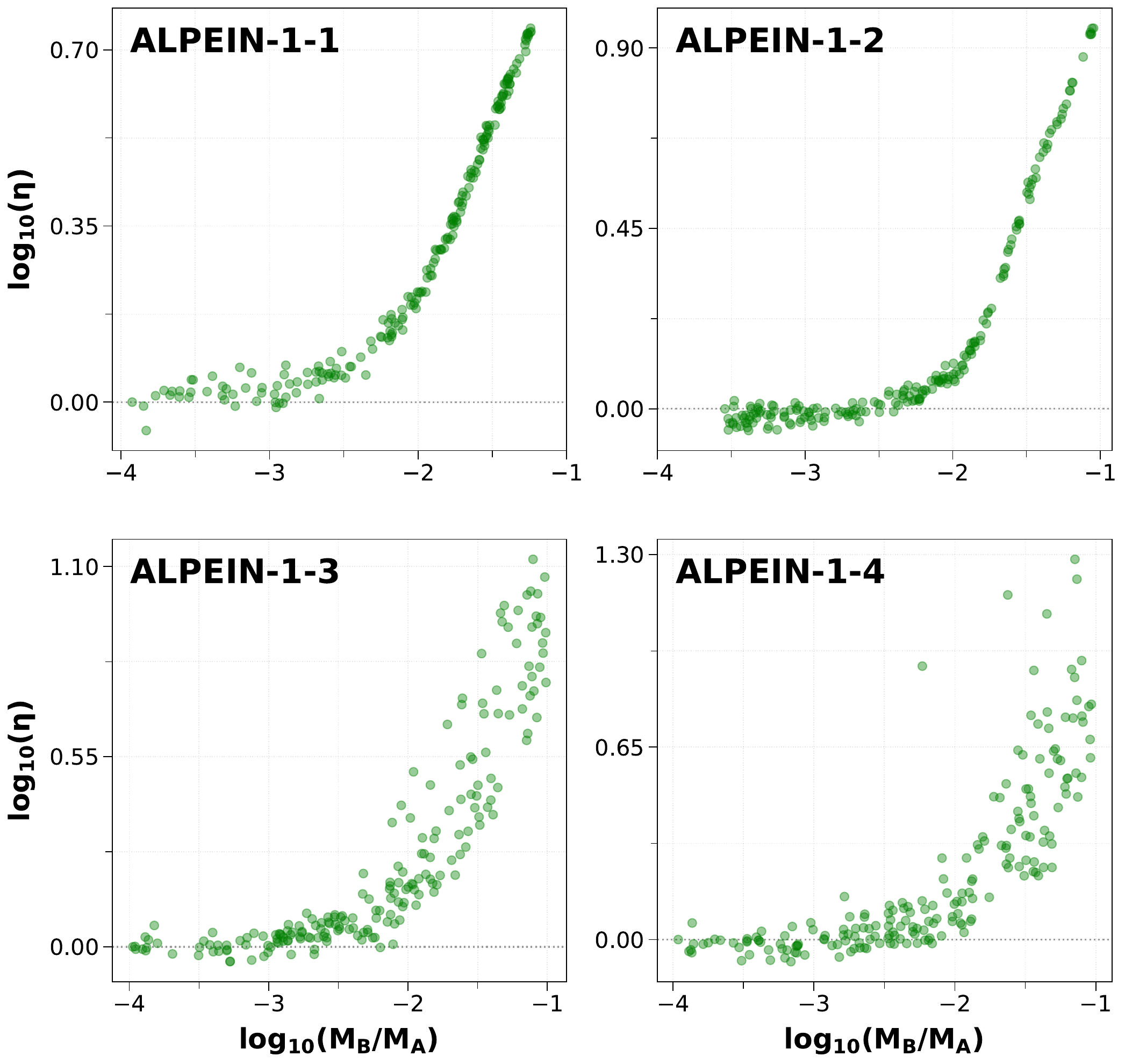}
    \caption{The results of our initial quad population comparisons with the ALPEIN-1 mass models. Each point corresponds to a comparison between an observed population of $N = 50$ quads and a mock population of $M = 2450$ quads. The similarity between these populations in our 3D space is quantified by the $\eta$ metric and is normalized with respect to the self-comparison $\eta_{\star}$ value. This 3-dimensional distance $\eta$ is plotted against the subhalo mass fraction within the galaxy the mock population of quads were obtained from. The observed population of quads for each ALPEIN-1 mass model was taken from a galaxy with a small subhalo mass fraction.}
    \label{fig:alp1-res}
    \vspace{-5pt}
\end{figure}

The $\eta$ metric is calculated by finding the distance between each observed quad and its $M/N$ closest mock quad neighbors, then averaging these distances. The exact process of calculating $\eta$ is as follows: The observed quads are randomly ordered in an array, i.e., $\vec\Theta_\textup{obs} = \{\boldsymbol\Theta_1, \boldsymbol\Theta_2, \ldots, \boldsymbol\Theta_N \}$. The distances between the first observed quad, $\boldsymbol\Theta_1$, and all $M$ mock quads is calculated. The shortest distance is recorded and the corresponding mock quad is removed from the population, leaving $M-1$ quads. The second observed quad $\boldsymbol\Theta_2$ is chosen and the process repeats, cycling through the set of observed quads $M/N$ times until each mock quad is mapped to a single observed quad. The recorded $M$ distances are then averaged to calculate $\eta$. To ensure that the ordering of the observed quads in our initial array does not affect $\eta$, we choose $M\!\gg\!N$. The metric can be expressed analytically and is given by
\begin{equation}
    \eta=\frac{1}{M}\sum_{i=1}^{M/N}\Big
    (\sum_{j=1}^{N}
    \Big|\mathbf\Theta_{\rm{obs},j}-\mathbf\Theta_{\rm{mock},ij}^*\Big| \Big)
\end{equation}
where $\vec\Theta_{\rm{mock},ij}^*$ is the $i$th closest mock quad in 3D space, that has not already been used, to the $j$th observed quad $\vec\Theta_{\rm{obs},j}$. The resulting $\eta$ value can be thought to be describing the quality of how well a mock population fits an observed population. The smaller the $\eta$, the more similar the two quad populations are.

The ranges of our three parameters $\theta_{23}$, $\Delta\theta_{23}$, and $d_4/d_1$ differ by about one to two orders of magnitude. If left un-normalized, then the average 3-dimensional distance $\eta$ will be skewed by the parameter with the largest range, $\theta_{23}$. When calculating $\eta$, the parameters $\Delta\theta_{23}$ and $d_4/d_1$ are multiplied by $\times 3$ and $\times 100$, respectively.

\begin{figure*}
    \centering
    \includegraphics[width=0.91\textwidth]{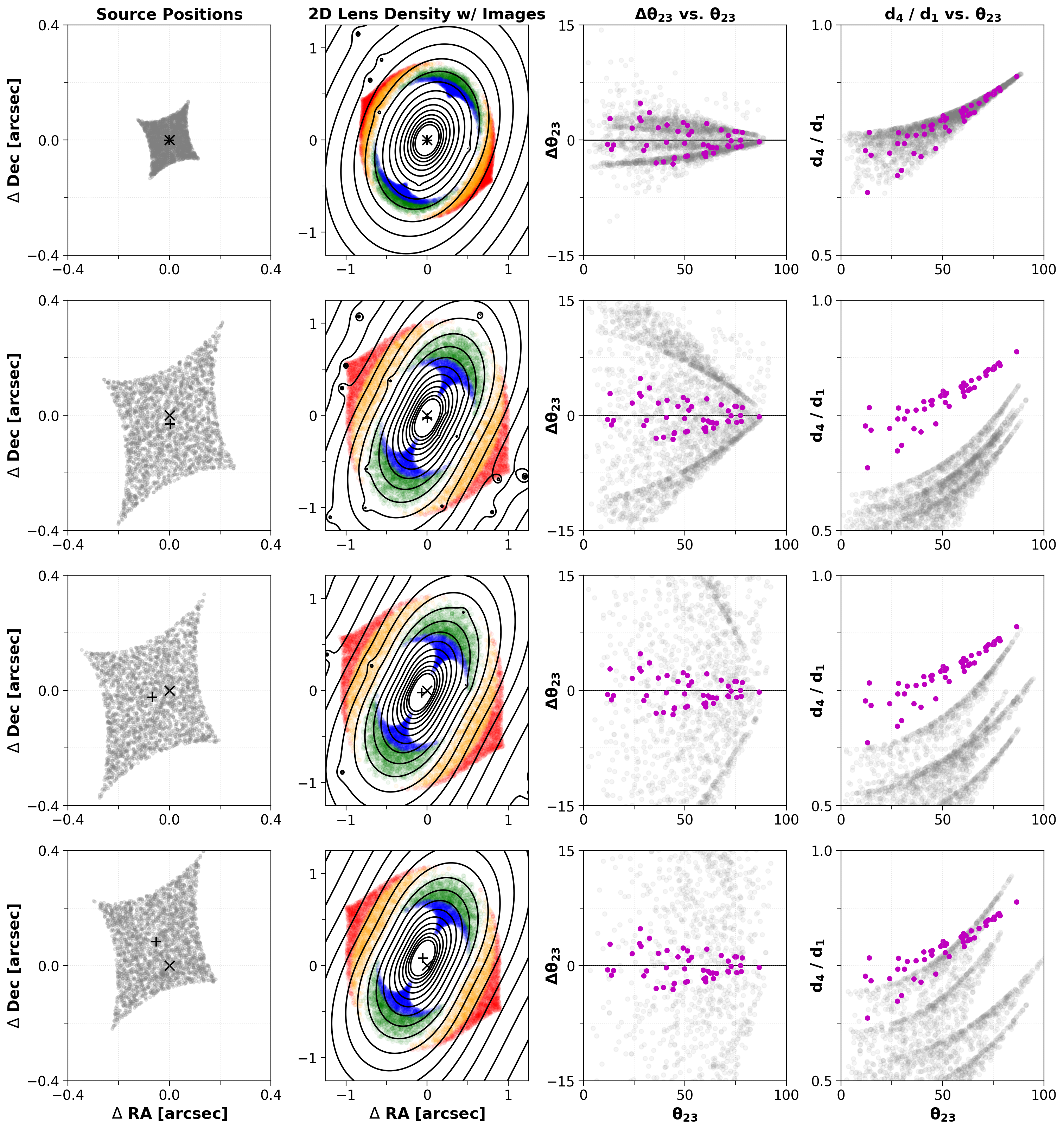}
    \caption{Examples of 4 ALPEIN-2 galaxy lenses (Rows 1-4). Column 1 shows randomly sampled source positions, where `+' is the center of the dark matter halo and `x' the center of the baryonic component. Column 2 shows the corresponding 2D lens density, with the first arriving images in red, second in yellow, third in green, and fourth in blue. Column 3 shows the comparison between the N=50 observed quads (magenta) taken from the first galaxy (Row 1) and M=2450 mock quads (gray) in $\Delta\theta_{23}$ vs. $\theta_{23}$. Column 4 shows the comparison between the same observed and mock quads, but this time in $d_4/d_1$ vs. $\theta_{23}$. The galaxies are arranged in order of increasing offset from top to bottom.}
    \label{fig:alp2}
\end{figure*}

As a preliminary test to see if our $\eta$ metric can distinguish between quad populations in our 3D space, we compare quad populations generated from individual galaxies (i.e., what we call galaxy-galaxy comparisons). For each ALPEIN-1 sub-model we choose the galaxy with the smallest subhalo mass fraction ($\log[\textup{M}_\textup{B} / \textup{M}_\textup{A}$] $\approx -4.0$) to form the observed population of $N = 50$ quads, which is randomly drawn from the galaxy's population of 2500 quads. A sub-model's observed population of quads, say for ALPEIN-1-1, is then compared to each of the 200 populations of $M = 2450$ mock quads generated from ALPEIN-1-1 galaxies, including the original population of quads the observed population was taken from. To ensure that our analysis is unbiased, we removed the 50 quads chosen to be the observed quad population from its population of 2500 quads and removed 50 random quads from all other galaxies. Therefore, each $\eta$ value will quantify the 3-dimensional average distance between $N = 50$ observed quads and $M = 2450$ mock quads. Columns 3 and 4 of Figure \ref{fig:alp1} show examples of how these comparisons work in the projections of our 3D space of observables, where the observed quad population (magenta) is taken from the first galaxy (Row 1).

\begin{table*}
    \centering
    \caption{The mass model parameters for the ALPEIN-2 model. The first six parameters are for the \texttt{alphapot} (Eq. \ref{eq:alphapot}) dark matter halo (`A'), the next four parameters are for the Einasto (Eq. \ref{eq:einasto}) baryonic component (`C'), and the last four parameters are for the Einasto dark matter subhalos (`B' \& `D'). The parameter M$^{\textup{lim}}_{\textup{BD}}$ describes the total upper mass limit for both lens-plane and LOS subhalos, and m$_\textup{min}$ is the smallest subhalo mass. The dark matter halo potential is chosen to be isothermal ($\alpha_\textup{A}$ = 1). The baryonic shape parameter, $\alpha_\textup{C}$, is consistent with the de Vaucouleurs profile ($n = 4 = 1 / \alpha$). And the dark matter subhalos' shape parameters, $\alpha_\textup{BD}$, are roughly consistent with the upper limit of shape parameters found in \citealt{2008springel}.}
    \begin{tabular}{cccccccccccc} \hline
        & \multicolumn{4}{c}{DM Halo ('A')} & \multicolumn{3}{c}{Baryonic Component ('C')} & \multicolumn{4}{c}{DM Subhalos ('B' \& 'D')}\\ \hline
        \multirow{2}{*}{ALPEIN-\#} & M$_\textup{A}$ & $s$ & $Q$ & $\alpha_\textup{A}$ & M$_\textup{C}$ & $r_s^\textup{C}$ & $\alpha_\textup{C}$ & M$^{\textup{lim}}_{\textup{BD}}$ & M$_\textup{B}^\textup{lim}$ / M$_\textup{D}^\textup{lim}$ & m$_\textup{min}$ & $\alpha_\textup{BD}$\\
        & [M$_\odot$] & [kpc] & & & [M$_\odot$] & [kpc] & & [M$_\odot$] & & [M$_\odot$] & \\ \hline
        ALPEIN-2 & $10^{11}$ & 0.275 & 0.68 $-$ 0.99 & 1.0 & $10^{10}$ & 1 $-$ 3 & 0.25 & $10^{10}$ & 0.25 $-$ 0.41$\bar{6}$ & $10^8$ & 0.25 \\ \hline
    \end{tabular}
    \label{tab:alpein2}
\end{table*}

If our $\eta$ metric is working, then we should find that two quad populations originating from galaxies with similar galaxy properties have a low $\eta$ value when compared to two populations with very different galaxies. Ideally the lowest $\eta$ value should belong to the self-comparison, designated by $\eta_{\star}$. This is the comparison between the observed population of quads and the population of quads it was originally removed from. In other words, our $\eta$ metric should `pick out' the galaxy the observed quads were taken from. (Fig. \ref{fig:etas} contains an illustration of this test, but for ALPEIN-2; see Section \ref{sec:alp2}). For our current test, because the observed quad population (for each respective sub-model) was taken from the galaxy with the smallest subhalo mass fraction, lower mass fractions should correspond to lower $\eta$ values and vice versa.

The $\eta$ comparison results for all four ALPEIN-1 models are shown in Figure \ref{fig:alp1-res}, where each $\eta$ is normalized by the self-comparison $\eta_{\star}$ value. Each point corresponds to one mock galaxy. Every ALPEIN-1 sub-model shows a clear correlation between $\eta$ and the subhalo mass fraction, with larger mass fractions correlating to larger $\eta$ values. The ALPEIN-1-1 and ALPEIN-1-2 models, which only vary one subhalo parameter, show strong correlations between the subhalo mass fraction and $\eta$. The dispersion in this correlation increases in ALPEIN-1-3, which has two free parameters, and in ALPEIN-1-4, which has three free parameters. The increase in dispersion is simply due to the two sub-models having more variations in properties. 

The correlation between the subhalo mass fraction and our $\eta$ metric in our simple model demonstrates a connection between our 3D space of lensing observables and one un-observable property of galaxy structure. Galaxies are much more complex in reality, however, which could affect the efficacy of our $\eta$ metric. Furthermore, we would like to know more about the inner structure of lensing galaxies beyond the subhalo mass fraction (i.e., dark matter halo ellipticity, baryonic ellipticity, offset, etc.). In the next section we stress test our $\eta$ metric with galaxy-galaxy comparisons of more realistic galaxies and determine if any correlations can be found between $\eta$ and galaxy model parameters.

\section{ALPEIN-2: Galaxy-Galaxy Comparisons} \label{sec:alp2}

The second rung of our analysis investigates galaxy-galaxy comparisons with quads generated from the ALPEIN-2 parametric galaxy model. ALPEIN-2 aims to be more realistic than ALPEIN-1 by having both an elliptical \texttt{alphapot} dark matter halo (`A') and a circular Einasto baryonic component (`C'). These two main mass components, A and C, are allowed to be displaced (or offset) up to 0.5 kpc away from each other to allow lopsidedness in the mass distribution. In addition to the dark matter halo and baryonic component, ALPEIN-2 has separate lens-plane (`B') and LOS (`D') subhalos modelled by Einasto profiles, which are now not restricted to three. The addition of these mass model complexities aims to produce more realistic galaxies. The exact properties of each are discussed in detail in Section \ref{sec:mods} and the values of components used in modelling are summarized in Table \ref{tab:alpein2}.

A population of 30 galaxies were created with the ALPEIN-2 mass model, with each galaxy having 2500 quads. Four example galaxies from ALPEIN-2 are shown in Figure \ref{fig:alp2}, Column 2, and are ordered from top to bottom with increasing central offset. The centers of the dark matter halo and baryonic components are denoted with `+' and `x', respectively. Note that despite the offset centers, each galaxy has a single mass peak. In the 2D projection of the FSQ, $\Delta\theta_{23}$ vs. $\theta_{23}$, galaxies typically create a distribution of quads that can be traced by a horizontal isosceles triangle centered on $\Delta\theta_{23} = 0$ (see Fig. \ref{fig:obs-dt23}). Increasing the mass component offset increases the base length of this triangle, widening the distribution of $\Delta\theta_{23}$ (see Fig. \ref{fig:alp2}, Column 3). As was found in \cite{2018gomer}, increasing centers' offset between the two main mass components results in substantial dispersion in $\Delta\theta_{23}$. The subhalo mass fraction does contribute to the dispersion and structure of $\Delta\theta_{23}$ (i.e., filaments), but its effect has been reduced from ALPEIN-1 to ALPEIN-2 due to individual subhalos now having more realistic, lower masses. Other mass model complexities do contribute as well, but the leading perturber in our 3D space of lensing observables is the magnitude of the central offset.

\begin{figure}
    \centering
    \includegraphics[width=1\columnwidth]{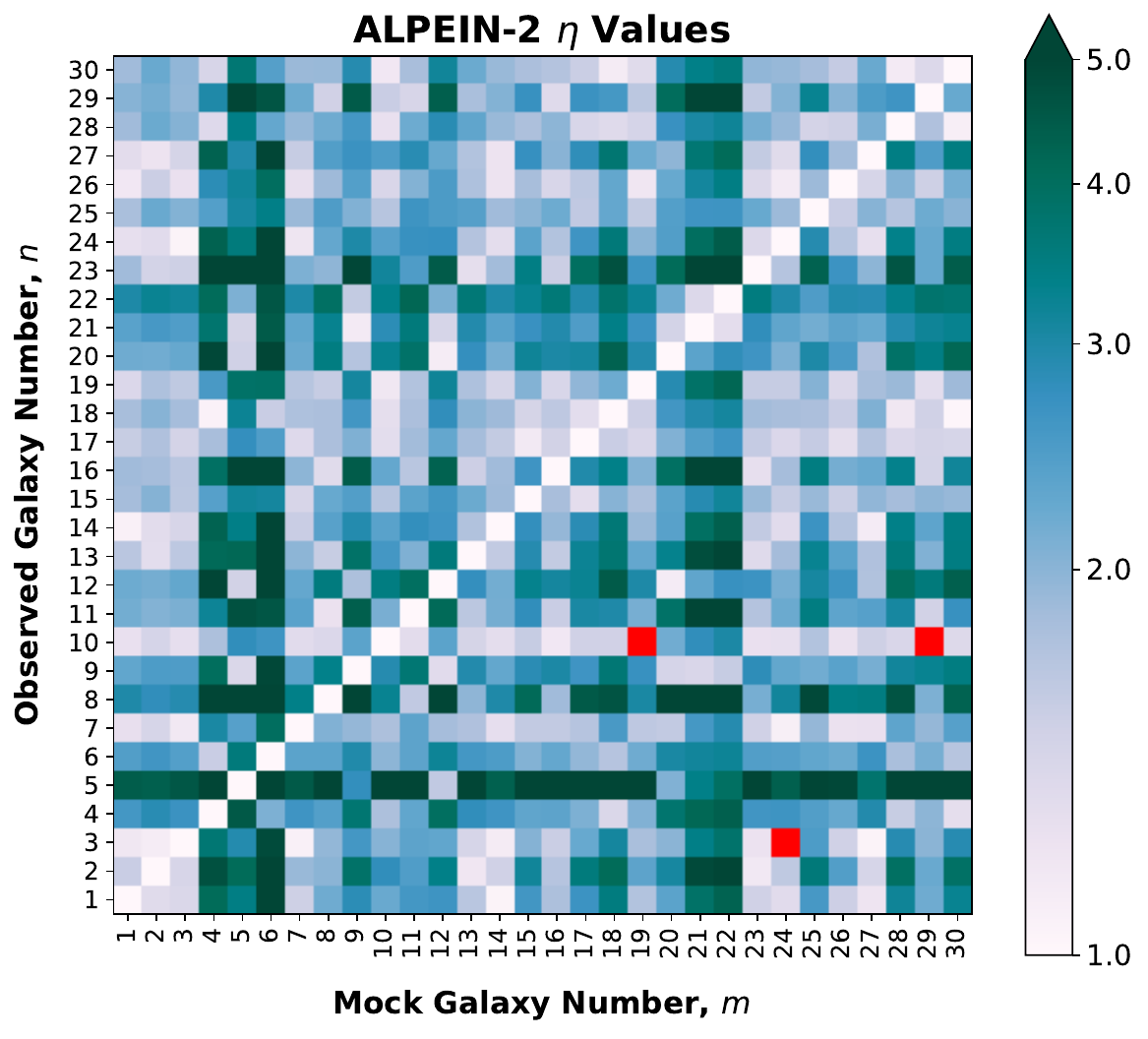}
    \caption{The results of comparing 50 observed quads from each of 30 lens galaxies plotted along the y-axis, to 2450 mock quads each from the same set of 30 lens galaxies plotted along the x-axis. The comparison metric is the average distance $\eta$. The self-comparison cases, where observed and mock quads come from the same lens galaxy, lie along the diagonal. The ordering of the 30 galaxies is random.}
    \label{fig:etas}
    \vspace{-0.5cm}
\end{figure}

In an effort to understand how offset is affecting $\Delta\theta_{23}$, we create three groups of galaxies: one where offsets vary only parallel to the main lens position angle (PA), another where offsets vary perpendicular to the main lens PA, and one with full azimuthal freedom. For each set of galaxies the magnitude of the lens galaxy's offset is compared to its quad populations' $\Delta\theta_{23}$ RMS deviation from zero. We find strong linear correlations between $\Delta\theta_{23}$ RMS and offset, and that perpendicular offsets with the same magnitude result in larger magnitudes of $\Delta\theta_{23}$ than parallel offsets. The results for azimuthal freedom yielded $\Delta\theta_{23}$ values between the two correlations (see Appendix \ref{sec:off}). The increasing dispersion of $\Delta\theta_{23}$ due to central offset with azimuthal freedom can be seen in Column 3 of Figure \ref{fig:alp2}, where galaxies are ordered by increasing offset.

The results of ALPEIN-1 demonstrated that our $\eta$ metric can differentiate between simple galaxies. To determine the effectiveness of $\eta$ with more realistic galaxies, we conduct galaxy-galaxy comparisons in our 3D space of observables with 30 galaxies generated from ALPEIN-2. Each of the 30 galaxies creates a $N = 50$ observed quad population, which is compared to every other galaxy's mock population of $M = 2450$ quads, including itself, resulting in 900 $\eta$ values. \footnote{As in ALPEIN-1, the $N = 50$ observed quads are removed from its galaxy's population of quads, and 50 random quads are set aside from every mock population of $M = 2500$ as to not bias our results.} The magnitude of the $\eta$ value quantifies the `likeness' of two different quad populations in our 3D space, and thus their lens galaxies. If our $\eta$ metric works, then we should find that the self-comparison, i.e., when the observed population of quads is compared to the population of quads it was taken from, is the lowest $\eta$ value for all comparisons. In other words, for each set of 30 comparisons we should recover the originating observed galaxy from our analysis. 

\begin{table*}
    \centering
    \caption{The mass model parameters for the ALPEIN-3 model. The first five parameters are for the \texttt{alphapot} (Eq. \ref{eq:alphapot}) dark matter halo and the next five parameters are for the \texttt{alphapot} baryonic component. The subhalo generation in ALPEIN-3 does not differ from that in ALPEIN-2. See Table \ref{tab:alpein2} for the parameters of subhalo generation. The dark matter potential is isothermal ($\alpha_\textup{A}$ = 1). The baryonic potential slope, $\alpha_\textup{C}$, is set to 0.7 because the baryonic component needs to dominate the center of the potential and then drop off faster than the dark matter component.}
    \begin{tabular}{ccccccccccc} \hline
        & \multicolumn{5}{c}{DM Halo (`A')} & \multicolumn{5}{c}{Baryonic component (`C')} \\ \hline
        \multirow{2}{*}{ALPEIN-\#} & M$_\textup{A}$ & $s_\textup{A}$ & $\theta_\textup{A}$ & $Q_\textup{A}$ & $\alpha_\textup{A}$ & 
        M$_\textup{C}$ & $s_\textup{C}$ & $\theta_\textup{C}$ & $Q_\textup{C}$ & $\alpha_\textup{C}$ \\
        & [M$_\odot$] & [kpc] & [deg] & & & [M$_\odot$] & [kpc] & [deg] & & \\ \hline
        ALPEIN-3 & $10^{11}$ & 0.275 & 30 & 0.68 $-$ 0.99 & 1.0 & $3 \times 10^{10}$ & 0.275 & $0-60$ & $0.48-1.19$ & 0.7 \\ \hline
    \end{tabular}
    \label{tab:alpein3}
\end{table*}

The resulting distribution of $\eta$ values from our ALPEIN-2 comparison of 30 galaxies are shown in Figure \ref{fig:etas}. Each of the galaxies is arbitrarily labelled from $1 - 30$. The color-coded $\eta$ value for a given cell is the result of comparing an observed $N = 50$ quad population taken from the $n$th galaxy, labeled on the y-axis, to $M = 2450$ quads taken from the $m$th mock galaxy, labeled on the x-axis. Each row corresponds to 30 comparisons with the $n$th observed galaxy to all $m$ mock galaxies. The self-comparison cases, where the observed and mock quads come from the same lens galaxy (i.e., $n = m$), lie along the diagonal. For these, the average distance $\eta$ should be the shortest (which we normalize to equal 1). If observed and mock quads are drawn from different galaxy lenses (off-diagonal cases), $\eta$ should be larger than 1, represented by darker colors in the logarithmic color scheme. Of the $30\times 30=900$ total comparisons, our $\eta$ metric works as intended for 897. Where it does not (3 red squares; $\eta\gtrsim 0.97$), the observed and mock galaxies happen to be very similar. 

The ability of our $\eta$ metric to `pick out' the galaxy the observed quad population originated from shows its efficacy at distinguishing between two quad populations in our 3D space of lensing observables. Furthermore, mock galaxies with properties similar to the observed galaxy were generally found to have lower $\eta$ values than mock galaxies with different properties. For example, the first three galaxies ($n,m$ = 1, 2, 3), which resulted in small $\eta$ values when compared to each other, have the same ellipticity and all have generally small offsets. On the other hand, the sixth galaxy ($n,m$ = 6), which resulted in $\eta \ge 5$ for all of the first three galaxies, differs from these galaxies the most in terms of ellipticity and has a large offset. The fact that comparing galaxies with properties similar to the observed galaxy results in low normalized $\eta$ values (1 $\le$ $\eta$ $\lessapprox$ 2), and high normalized $\eta$ values ($\eta \gtrapprox 2$) for dissimilar galaxies, demonstrates that there exists a mapping between galaxy parameters and our 3D space of quad image observables.

\section{ALPEIN-3: Population-Population Comparisons} \label{sec:alp3}

\begin{figure*}
    \centering\
    \includegraphics[width=0.9\textwidth]{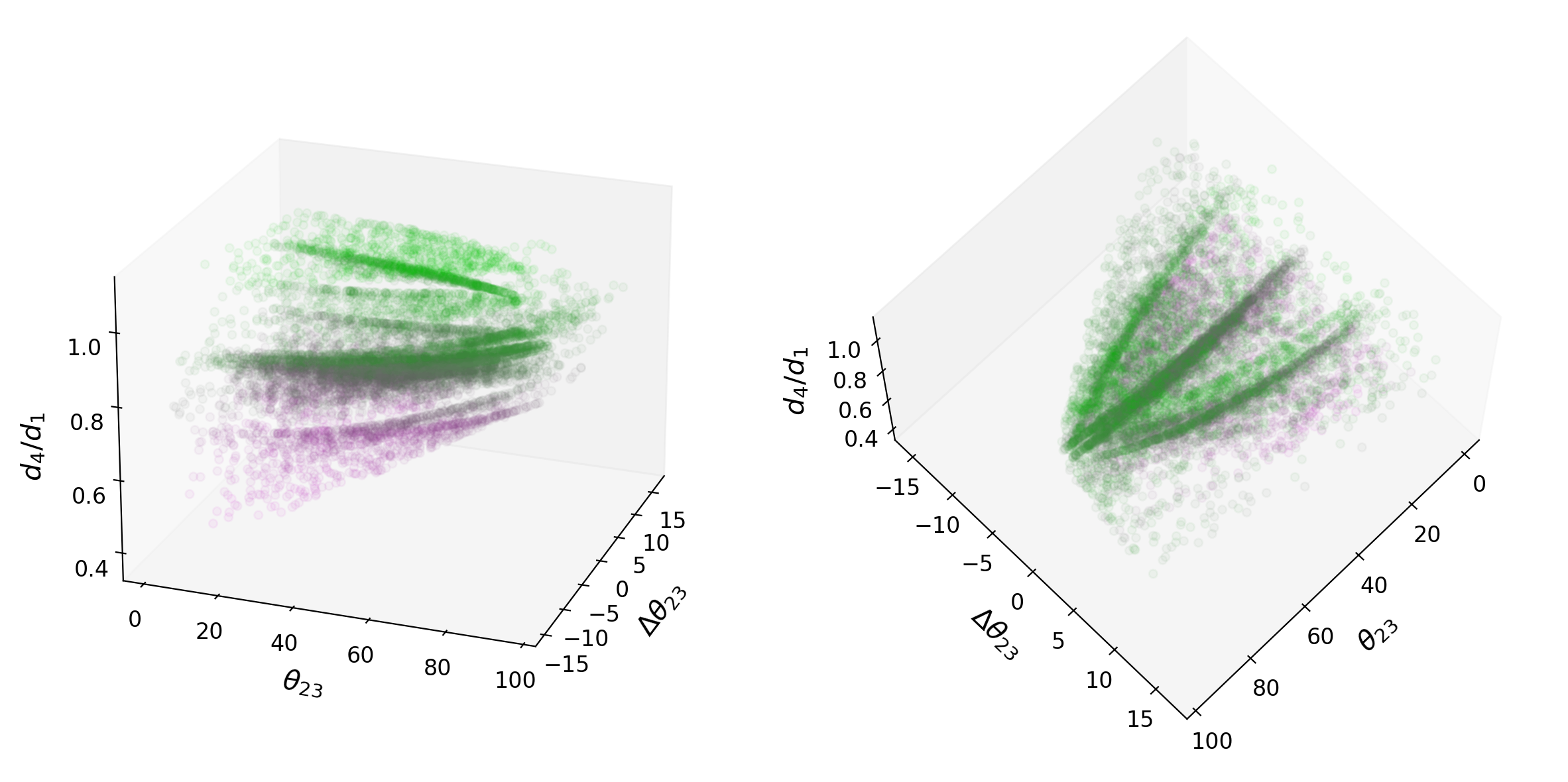}
    \includegraphics[width=0.9\textwidth]{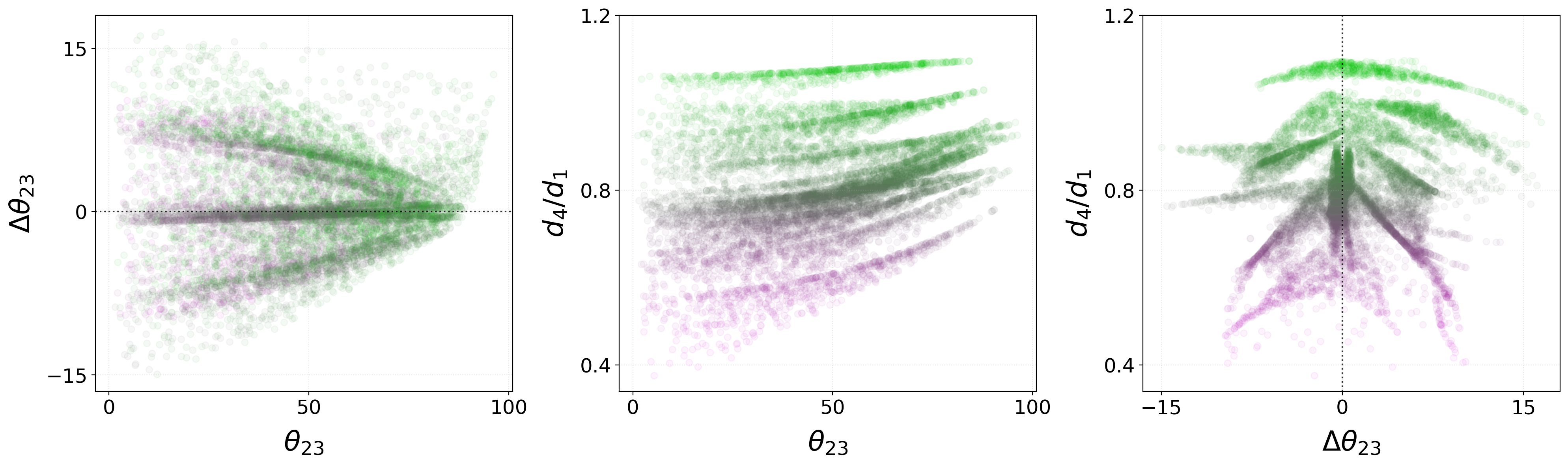}
    \caption{Our 3D space of lensing observables (top row) and the three 2D projections of the space (bottom row). Plotted are 10,000 quads from four ALPEIN-3 galaxies, where each point corresponds to one quad. The points are color-coded based on $d_4/d_1$: the smaller $d_4/d_1$ values are magenta and larger $d_4/d_1$ values are green. The first 2D projection of $\Delta\theta_{23}$ vs. $\theta_{23}$ (bottom row; left) is the 2D projection of the FSQ and can be seen in Column 3 of Figures \ref{fig:alp1} and \ref{fig:alp2} for ALPEIN-1 and ALPEIN-2, respectively. Figure \ref{fig:obs-dt23} shows the same 2D projection, but for the observed quads. The second 2D projection of $d_4/d_1$ vs. $\theta_{23}$ can be seen in Column 4 of Figures \ref{fig:alp1} and \ref{fig:alp2}}.
    \label{fig:3d-space}
\end{figure*}

The ALPEIN-3 parametric galaxy model differs from the ALPEIN-2 model in two ways: the baryonic component is now modelled by an \texttt{alphapot} potential, allowing it to have non-zero ellipticity and position angle, and the baryonic mass is increased to $3 \times 10^{10}$ M$_{\odot}$. The axial ratio $Q_\textup{C}$ and position angle $\theta_\textup{C}$ of the baryonic component are allowed to vary from the main dark matter halo values by $Q_\textup{A} \pm 0.2$ and $\theta_\textup{A} \pm 30\degree$, creating further asymmetry in the mass distribution. The parameter values for ALPEIN-3 are summarized in Table \ref{tab:alpein3}. Subhalo generation in ALPEIN-3 does not differ from that in ALPEIN-2. A total of 1,200 galaxies were created with 2,500 quads taken from each galaxy, resulting in 3 million total quads. Figure \ref{fig:3d-space} shows a total of 10,000 quads taken from four ALPEIN-3 galaxies in our 3D space of lensing observables, $\theta_{23}$, $\Delta\theta_{23}$, and $d_4/d_1$, and in the three 2D projections of the 3-dimensional space. There exists distinctive filamentary structure in each 2-dimensional projection, which greatly differs depending on what galaxy or galaxies the quad population is taken from. Because different galaxy properties are mapped to different regions of our 3D space, these unique quad configurations demonstrate why our $\eta$ metric can resolve between quad populations.

The ALPEIN-3 data set consist of quad populations from a population of galaxies, akin to the true observed quad population; we use these for population-population comparisons. In the previous galaxy-galaxy comparisons (Sections~\ref{sec:alp1} and \ref{sec:alp2}), the similarity or dissimilarity between two galaxies was quantified by the difference between one galaxy parameter, like subhalo mass fraction, which only had one value. However, with quad populations arising from a population of galaxies in ALPEIN-3, we instead compare distributions of values for each galaxy parameter. For simplicity we choose to represent these parameter distributions as Gaussian. In reality, the true observed distribution of a parameter could take any form.

A population of galaxies can be represented by a point in the $k$-dimensional space of galaxy parameters, $\mathbf{Z} = (\overline{Q}_\textup{A}, \overline{Q}_\textup{C}, ...)$, where $\overline{Q}_\textup{A}$ is the mean of the Gaussian distribution of $Q_\textup{A}$, etc., and $k$ denotes the number of constrained parameters. If the standard deviation of a Gaussian distribution is kept constant between the two populations, then the $k$-dimensional distance between the means of the observed parameter distributions, $\mathbf{Z}_\textup{obs}$, and the mock parameter distributions, $\mathbf{Z}_\textup{mock}$, is given by
\begin{equation} \label{eq:dist}
    \zeta ~=  |\mathbf{Z}_{\textup{obs}} - \mathbf{Z}_{\textup{mock}}| ,
\end{equation}
and can be compared with $\eta$. Each parameter, and thus the resulting $k$-dimensional distance $\zeta$, is normalized based on the range of allowed values for that parameter, resulting in $\zeta \in [0, 1]$. Galaxy populations with similar distributions of galaxy parameters, and thus similar Gaussian means, will have a small $\zeta$ value (i.e., $\zeta \approx 0$). Quad populations arising from these similar galaxy populations should result in correspondingly low $\eta$ values when compared to each other. Galaxy populations with dissimilar parameter distributions, will have a large $\zeta$ value (i.e., $0 \lessapprox \zeta \leq 1$), and should conversely result in large $\eta$ value between their quad populations. Utilizing this framework, we test the efficacy of our $\eta$ metric with population-population comparisons.

For simplicity, we limit our current analysis to the three primary galaxy parameters which create the leading galaxy asymmetries: the dark matter ellipticity $Q_\textup{A}$, baryonic ellipticity $Q_\textup{C}$, and central offset. By constraining only one ($\mathbf{Z} \subseteq \mathbb{R}^1$) or two ($\mathbf{Z} \subseteq \mathbb{R}^2$) of these galaxy parameters we create galaxy populations in five different ways: $\mathbf{Z_1} = (\overline{Q}_\textup{A})$, $\mathbf{Z_2} = (\overline{Q}_\textup{C})$, $\mathbf{Z_3} = (\overline{\textup{Offset}})$, $\mathbf{Z_4} = (\overline{Q}_\textup{A}, \overline{Q}_\textup{C})$, and $\mathbf{Z_5} = (\overline{Q}_\textup{A}, \overline{\textup{Offset}})$. Populations created by constraining one galaxy parameter (e.g., $\mathbf{Z_1}$) are labelled as single-parameter populations and populations created by constraining two parameters (e.g., $\mathbf{Z_4}$) are labelled dual-parameter populations. All other non-constrained galaxy parameters are left to be random.

The three main parameters that one generally associates with massive galaxies are the ones defining the Fundamental Plane of ellipticals (FP). However, the velocity dispersion, effective radius, and surface brightness all describe the mass normalization of the galaxy and its radial extent, implicitly assuming circular symmetry. Our lensing analysis is designed to probe the azimuthal asymmetries of lens galaxies, so our primary parameters are different from those defining the FP. The distribution of galaxies in the FP is well known, while our analysis will provide complementary information. Just like parameters of the FP are correlated, the parameters describing azimuthal asymmetries are also likely to be correlated with each other \citep[e.g.,][]{2019shajib}. 

We explore the two extremes of azimuthal parameter interdependence: perfectly correlated, and completely uncorrelated. In $\mathbf{Z_4} = (\overline{Q}_\textup{A}, \overline{Q}_\textup{C})$ the two galaxy parameters are fully correlated: $\overline{Q}_\textup{A} = \overline{Q}_\textup{C}$ for each $\mathbf{Z_4}$. In $\mathbf{Z_5} = (\overline{Q}_\textup{A}, \overline{\textup{Offset}})$ the two parameters are left uncorrelated.

Three types of quad populations are generated for each parameter selection $\mathbf{Z_1}$-$\mathbf{Z_5}$: observed, mock, and self-comparison. Each has a total of 10 observed populations of $N = 60$ quads and 600 mock populations of $M = 3600$ quads. These mock quad populations are taken from galaxy populations with the constrained parameter mean(s) randomized within our priors (See Table \ref{tab:alpein3}). Additionally, to create a baseline for the $\eta$ metric, we take $M = 3600$ quads from each observed galaxy population to create 10 self-comparison quad populations. In total, each of $\mathbf{Z_1}$-$\mathbf{Z_5}$ has 10 observed, 600 mock, and 10 self-comparison quad populations. As mentioned earlier, the standard deviation for the constrained galaxy parameters are constant and equal between all observed, mock, and self-comparison galaxy populations.

\begin{figure*}
    \centering
    \includegraphics[width=0.9\textwidth]{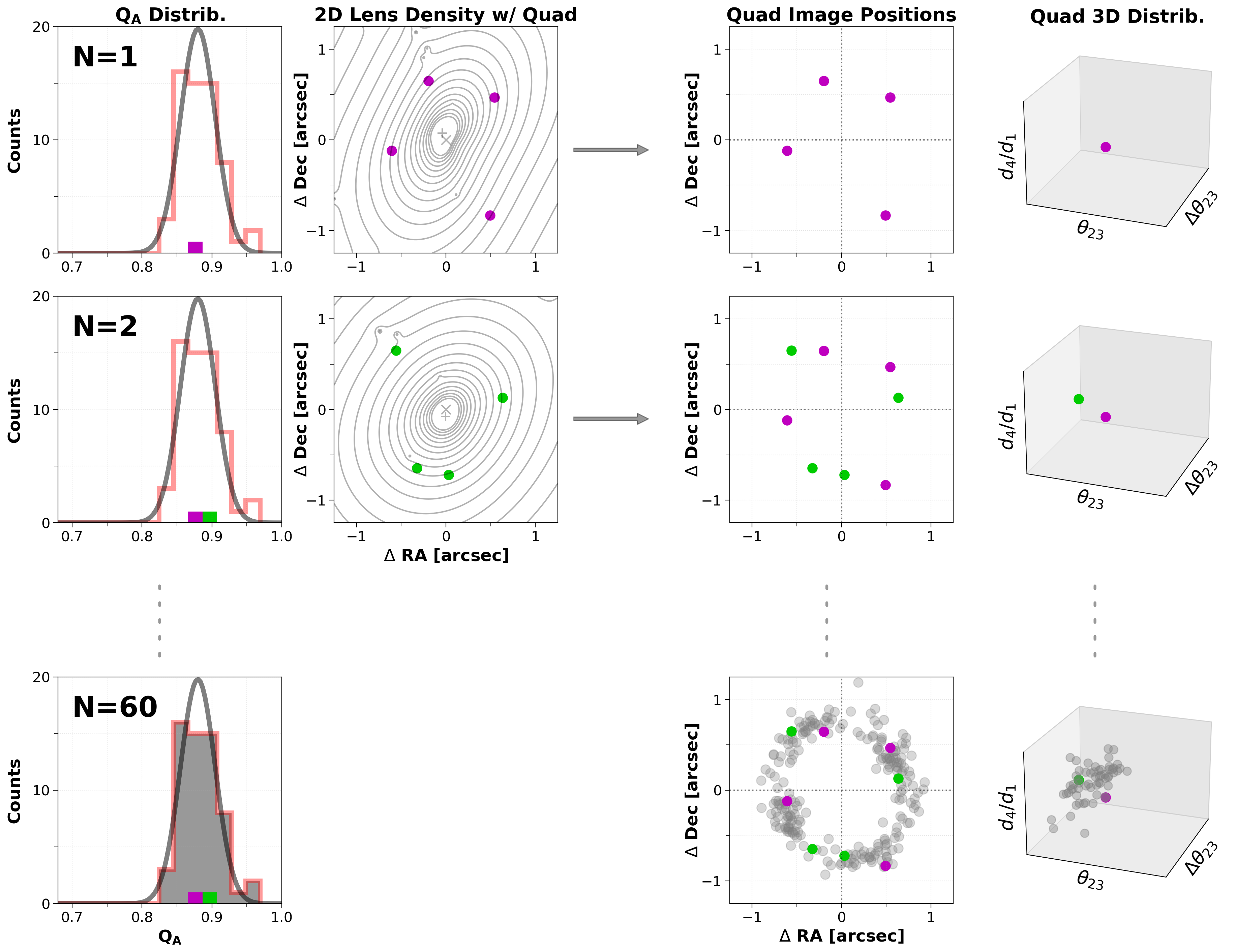}
    \caption{The process for creating an observed quad population from a population of galaxies for $\mathbf{Z_1} = (\overline{Q}_\textup{A})$. Column 1 shows the distribution of dark matter ellipticity $Q_\textup{A}$ as galaxies are added to the observed population ($N = 0 \rightarrow 60$). The distribution is defined by a 1D Gaussian (smooth gray curve) and the red Gaussian histogram is one realization of the gray target distribution. Column 2 shows the 2D lens density (convergence) for the galaxy corresponding to the $Q_\textup{A}$ value added in Column 1. One of the galaxy's quads is over-plot, magenta for $N = 1$ and green for $N = 2$. These quads are collected and over-plot without a convergence map in Column 3 and in our 3D space of observables in Column 4. The bottom row shows the final observed distribution of $Q_\textup{A}$ and observed quad population. Mock populations are created in the same manner, but 60 quads are taken from each of the $N = 60$ galaxies instead of one.}
    \label{fig:mega}
\end{figure*}

The process of creating an observed quad population for $\mathbf{Z_1}$ is shown in Figure \ref{fig:mega} and is as follows: A Gaussian distribution for the galaxy parameter $Q_\textup{A}$ is defined by an input mean and standard deviation (Column 1; gray curve). Galaxies are then randomly selected from the supply of 1,200 ALPEIN-3 galaxies and added to the observed galaxy population until a good approximation of the (gray) Gaussian distribution is obtained (red histogram; one realization). The bottom panel of Column 1 shows the final observed distribution of $Q_\textup{A}$. If a galaxy has been chosen for the observed galaxy population, one of its 2,500 quads is randomly selected and placed into the observed quad population (Columns 2 and 3). This quad corresponds to one point in our 3D space of lensing observables (Column 4). Row 1 left-to-right shows this process for the first $N = 1$ galaxy. The final observed galaxy distribution of $N = 60$ dark matter ellipticities, $Q_\textup{A}$, and the observed population of $N = 60$ quads in the 3D space is shown in Row 3. The methodology of creating dual-parameter populations is the same procedure as single-parameter populations, but instead uses two galaxy parameters. For example, when creating an observed population for $\mathbf{Z_4}$, a galaxy will only be added to the population if both its $Q_\textup{A}$ and $Q_\textup{C}$ values are defined within the chosen distributions: a galaxy must occupy both parameter spaces at once. If one parameter lies outside the desired distribution, or if the histogram bin corresponding to the value is already filled (see Fig. \ref{fig:mega}; red histogram), then the galaxy is rejected and a new one selected. Mock and self-comparison quad populations are generated in the same manner, but a total of 60 quads are taken from each galaxy instead of one quad per galaxy (Columns 2 and 3).

\begin{figure*}
    \centering
    \includegraphics[width=0.9\textwidth]{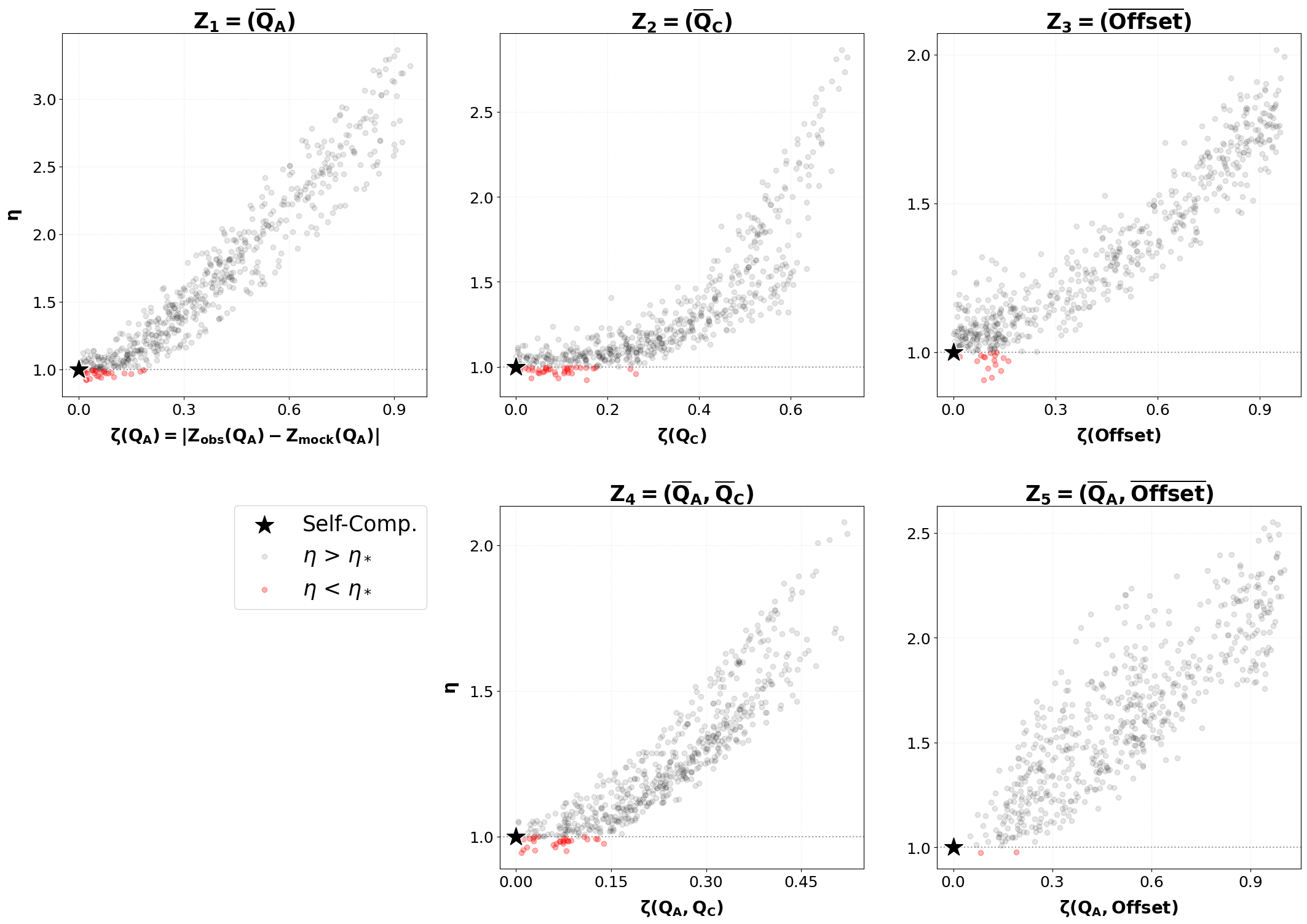}
    \caption{The results of the ALPEIN-3 population-population comparisons for each of the five parameter selections, $\mathbf{Z_1}$-$\mathbf{Z_5}$. Each point corresponds to a population-population comparison with 3-dimensional average distance $\eta$ and distance between the observed and mock Gaussian parameter means (titles) $\Delta\eta$. All of the populations in Row 1 are single-parameter populations, which are defined by having only one galaxy parameter defined by a Gaussian ($\mathbf{Z} \subseteq \mathbb{R}^1$), whereas Row 2 are dual-parameter populations with two Gaussian distributions {($\mathbf{Z} \subseteq \mathbb{R}^2$)}. The distributions of the galaxy parameters other than the constrained parameter(s) are left to be random. }
    \label{fig:alp3-res}
\end{figure*}

Each observed quad population is compared to 60 mock populations and its self-comparison population in our 3D space of lensing observables. Therefore, each observed population results in 61 $\eta$ values, which are normalized based on the self-comparison $\eta$ value, $\eta_{\star}$. Because each of $\mathbf{Z_1}$-$\mathbf{Z_5}$ contains 10 observed quad populations, each one has a total of 610 $\eta$ values after comparisons are completed. The similarity between the observed and mock galaxy populations is estimated by finding the normalized distance between the observed and mock means of the selected parameter(s) $\zeta$ (Eq. \ref{eq:dist}). Similarly to how $\eta$ is the distance between two quad populations in our 3D space of lensing observables, $\zeta$ is the distance between two galaxy populations in the $k$-dimensional space of galaxy properties, where $k$ is the number of constrained parameters. The key feature that our method relies on is that the distance in one space should be closely related the distance in the other. Next, we demonstrate that this is, in fact, the case.

The ALPEIN-3 population-population comparison results are shown in Figure \ref{fig:alp3-res}, where each graph corresponds to galaxy populations created with one of the five parameter selections: $\mathbf{Z_1} = (\overline{Q}_\textup{A})$, $\mathbf{Z_2} = (\overline{Q}_\textup{C})$, $\mathbf{Z_3} = (\overline{\textup{Offset}})$, $\mathbf{Z_4} = (\overline{Q}_\textup{A}, \overline{Q}_\textup{C})$, and $\mathbf{Z_5} = (\overline{Q}_\textup{A}, \overline{\textup{Offset}})$. The $\eta$ values are normalized by the self-comparison $\eta_{\star}$ and plotted against the $k$-dimensional, normalized distance between the Gaussian parameter means $\zeta$. Each point corresponds to one population-population comparison, and each graph shows 10 realizations of comparing an observed population of quads to its 61 mock quad populations. The self-comparison is defined to be located at ($\zeta$, $\eta$) = (0, 1), with 10 over-plotted points corresponding to the 10 observed populations. Population-population comparisons with $\eta > 1$ are colored grey and those with $\eta < 1$ are colored red.

All of the population-population comparisons resulted in correlations between our metric $\eta$ and $\zeta$. In general, larger distances between parameter means resulted in larger $\eta$ values with moderate dispersion. $\mathbf{Z_1}$, $\mathbf{Z_3}$, and $\mathbf{Z_5}$ have mock galaxy populations that span the full parameter space for its given parameter(s) (i.e., $\zeta \in [0, 1)$). In the other two parameter selections, $\mathbf{Z_2} = (\overline{Q}_\textup{C})$ and $\mathbf{Z_4} = (\overline{Q}_\textup{A}, \overline{Q}_\textup{C})$, mock populations do not span the full prior space (primarily in $Q_\textup{C}$), which is simply due to the limited pool of ALPEIN-3 galaxies. The dispersion in $\eta$ for the dual-parameter comparisons (two lower panels of Fig. \ref{fig:alp3-res}) are larger than in the single-parameter comparisons, and could be due to the galaxy parameter space, $\mathbf{Z}$, being larger. That is, the dual-parameter comparisons increase the dimensionality of $\mathbf{Z}$ and thus increase the range of directionallities of $\zeta$. 

In addition to the correlations, the region around the self-comparison is also of interest.  
All five $\mathbf{Z_1}$-$\mathbf{Z_5}$ comparisons in Figure \ref{fig:alp3-res} found a handful of $\eta$ values that are less than the self-comparison (red points). That is not to say that our $\eta$ value is not effective at distinguishing two populations of quads originating from galaxy populations, but there exists some uncertainty in our metric when the respective parameter means are similar (i.e. $\zeta$ is small). Of the five parameter selections, $\mathbf{Z_2} = (\overline{Q}_\textup{C})$ was the least robustly estimated. Of the 600 mock populations $\approx 7\%$ had $\eta < 1$. Both $\mathbf{Z_1} = (\overline{Q}_\textup{A})$ and $\mathbf{Z_4} = (\overline{Q}_\textup{A}, \overline{Q}_\textup{C})$ were similarly well estimated with only $\approx 5\%$ of comparisons resulting in $\eta < 1$. $\mathbf{Z_3} = (\overline{\textup{Offset}})$ was the best estimated single-parameter with only $\approx 3\%$, and $\mathbf{Z_5} = (\overline{Q}_\textup{A}, \overline{\textup{Offset}})$ being the best estimated overall with only 2 mocks, or $\approx$ 0.3\%, below unity. The dual-parameter combination of dark matter ellipticity, $Q_A$, and central offset improves on their single-parameter counterparts by reducing the number of comparisons with $\eta < 1$. The dispersion of $\eta$ values in the region near the self-comparison is likely due to other galaxy properties that are not taken into account by $\zeta$, like the angle between $Q_A$ and $Q_C$, which are affecting $\eta$.

We foresee two effects as the dimensionality of $\mathbf{Z}$, and thus $\zeta$, increases to include the full parametric galaxy property-space (i.e., $k \rightarrow$ many): the increase in the correlation's dispersion away from the origin, $(\zeta, \eta) = (0, 1)$, and the decrease of points in the immediate region of the self-comparison. The dimensionality of $\mathbf{Z}$ increases the number of ways any given $\zeta$ value can be created. Therefore, the number of comparisons giving rise to one $\zeta$ value will increase, and thus increase the dispersion in $\eta$ at larger $\eta$ values. Regarding the second effect, if all galaxy properties are taken into account by $\mathbf{Z}$, then there should be no remaining galaxy property that can induce dispersion in $\eta$ when $\zeta = 0$ and the dispersion near the self-comparison should decrease. A preliminary illustration of this result can be seen in the third column of Figure \ref{fig:alp3-res}, where the bottom panel includes both leading perturbers Q$_A$ and offset. Increasing the dimensionality of galaxy population parameters and $\zeta$ should improve our ability to find the true galaxy properties.

The correlations shown in Figure \ref{fig:alp3-res} demonstrate not only that specific distributions of galaxy properties are mapped to our 3D space of lensing observables, but also demonstrates the effectiveness of our $\eta$ metric. Our results pave a clear path towards comparisons with the true observed quad population.

\section{Summary \& Future Work} \label{sec:sum}

In the near future, surveys like the Rubin Observatory's LSST and Euclid will deliver an unprecedented number of galaxy-scale lensed quasars and supernovae, with well-defined selection criteria. These large, uniform data sets allow one to re-imagine how to approach strong galaxy lensing. Our analysis, described in detail in Sections~\ref{sec:alp1}-\ref{sec:alp3} and summarized below, deviates from the conventional approach by modelling galaxy populations and allowing for possible correlations between galaxy properties, without mass modelling of individual lenses. We use quads only, because they have $3\times$ as many image constraints as doubles do. From the observed quad population we extract model-independent image properties, $\mathbf\Theta$ = ($\theta_{23}$, $\Delta\theta_{23}$, $d_4/d_1$), and compare these to those extracted from many mock galaxy populations. We use point sources only because these allow us to measure $\mathbf\Theta$ unambiguously and accurately.

The goal of our analysis in this paper is to establish a framework for comparing mock populations of quads to the observed quad population in order to constrain the true distributions of galaxy population parameters. To quantify the difference between two populations of quads in our 3-dimensional space of lensing observables, $\mathbf\Theta$, we create the metric $\eta$, which is an average 3-dimensional distance between the observed and mock quads (see Section \ref{sec:eta} for a more precise explanation). When two quad populations are compared, $\eta$ will approach zero the more similar the populations are in our 3D space. With our choice of image observables, $\mathbf\Theta$, $\eta$ mainly quantifies the difference in azimuthal asymmetry between the galaxy populations the quads were extracted from. This difference can be due to altering one galaxy parameter (Section \ref{sec:alp1}), or altering many parameters at once (Sections \ref{sec:alp2} and \ref{sec:alp3}).

The main finding of this paper is that the distance $\eta$ between two quad population in the 3D space of image properties, and the distance $\zeta$ between the two corresponding galaxy-lens populations in the space of galaxy population properties, $\mathbf{Z}$, are well correlated. This is illustrated in Figure~\ref{fig:alp1-res} for ALPEIN-1, and in Figure~\ref{fig:alp3-res} for ALPEIN-3. In the case of ALPEIN-1, the galaxy property we tested was the subhalo mass fraction, while in the case of ALPEIN-3, the galaxy population properties tested were ellipticities of the dark matter and baryonic distributions, $Q_A$, and $Q_C$, and the offsets of their centers.

The true observed quad population originates from a population of galaxies (one quad per galaxy) and not from a single galaxy. So, in ALPEIN-3 (Section \ref{sec:alp3}), instead of galaxy properties being defined by singular values, for example $Q_\textup{A}$, like in galaxy-galaxy comparisons, they are instead defined by distributions at the population level. The simplest case is where only one galaxy property is given by a Gaussian distribution (i.e. $\mathbf{Z}$ one dimensional). All other galaxy properties are left to be random. The similarity of two galaxy populations is quantified by the distance between the observed and mock chosen parameter means, given by $\zeta$ (see Eq. \ref{eq:dist}). From our population-population comparisons, we find that the 3D distance between two quad populations, $\eta$, correlates with the multi-parameter distance between the two lensing galaxy populations, $\zeta$. In other words, the distribution of lens observables in our 3D space encodes the distribution of galaxy parameters in the multi-dimensional space of galaxy properties. Within some error, the galaxy population properties can be elucidated by the distribution of its quads in the 3-dimensional space of lensing observables.

With a perfect metric and infinitely large quad sets, a mock quad population taken from the same galaxy population as the observed one should have the lowest $\eta$ value. However, from our results in Figures \ref{fig:alp2} and \ref{fig:alp3-res}, we see that there exists some uncertainty in $\eta$. Picking the population-population comparison with the lowest $\eta$ value from any five of our ALPEIN-3 comparisons does not return the self-comparison, but instead returns a galaxy population with slightly inaccurate parameters. For example, the single-parameter $Q_\textup{A}$ comparisons have the lowest $\eta$ value corresponding to $\zeta(Q_\textup{A}) = 0.01$. Therefore, our future analysis will need to estimate errors on the derived population-level galaxy properties. 

The uncertainty in $\eta$ itself, which mostly arises due to the observed population being small, can be determined with many self-comparisons. The confidence levels (i.e., 1$\sigma$, 2$\sigma$, etc.) in $\eta$ will then be directly translated into the corresponding confidence levels in population-level galaxy parameters. Astrometric uncertainties in the observed population will be taken into account in a similar manner as \citealt{2018gomer} and our Figure \ref{fig:obs-dt23}. Because the goal of the current paper is to simply show the relation between $\eta$ and $\zeta$ this precise examination of the uncertainty in the derived galaxy properties will be investigated in the following paper when the final analysis has been established.

The goal of the future paper is to search the full multi-dimensional space of galaxy properties based on our parametric galaxy model. Exploring the parameter space in a more realistic setting can be achieved in a couple of ways depending on its size. The downhill simplex algorithm is sufficient for lower dimensionality spaces, whereas either Markov Chain Monte Carlo (MCMC) or a genetic algorithm (GA) are appropriate for a higher number of dimensions. Regardless of the chosen minimization algorithm, the parameter space will be explored using the $\eta$ metric. Common strong lensing selection biases will be applied to our mock quad populations to match the biased sample of galaxies probed by strong lensing \citep{2012arnseon,2018gomer,2023sonnenfeld}. 

Because we are currently not including external shear to account for environmental effects, using the current mock quad populations would result in biased ellipticities. Therefore, our future analysis will incorporate environmental effects from nearby and distant massive objects by including shear with magnitudes consistent with weak lensing literature \citep{1997keeton,2005dalal,2024etherington}. Additionally, our analysis will expand to include a wider range of parametric galaxy models.

In this paper, the $\eta$ metric is defined to be a 3-dimensional average distance, but it can be expanded to include any number of dimensions of quad image properties. Now that the efficacy of our analysis has been demonstrated, the full 6D parameter space of three relative angles and three distance ratios will be utilized for our subsequent analysis. A 7th dimension can be included as well, representing the size of the Einstein radius, $\theta_E$ of the quad. The discriminating power of $\eta$ with only three observables provides compelling results, thus we anticipate $\eta$'s resolution to improve with all six or seven dimensions.

\section{Conclusion} \label{sec:conc}

In this paper we present a novel analysis of quad populations in a 3D space of lensing observables $\boldsymbol\Theta$ = ($\theta_{23}$, $\Delta\theta_{23}$, and $d_4/d_1$), with the goal of extracting properties of the population of lensing galaxies. In contrast to standard individual mass modelling, we conduct forward modelling and do not perform any mass model fitting. The framework of our model-free analysis depends on the robustness of a metric to quantify the proximity of two quad populations in our 3D space of observables. We create and assess the metric, $\eta$, which is the average 3-dimensional distance between each quad and its closest neighbors. In a similar way, the proximity of two galaxy populations in the multi-dimensional space of galaxy properties can be quantified by the distance between the means of these parameter distributions, or $\zeta$. When comparing quad populations taken from a population of galaxies (population-population comparisons), we find the two distances $\eta$ and $\zeta$ to be correlated. That is, if the property distributions of the two galaxy populations are similar, and thus $\zeta$ is small, then their two quad populations should approximate each other in our 3-dimensional space of lensing observables, and $\eta$ should also be small. Therefore, the distribution of galaxy properties can be estimated from our $\eta$ metric.

Having established the efficacy of our $\eta$ metric and the potential of our analysis, we aim to extract the observed galaxy population properties from the population of observed quads in a follow-up paper. The current observed population of quads contains $N \approx 60$ quads and is expected to increase by orders of magnitude with the deluge of lensed quasars and supernovae from large scale surveys like the Rubin Observatory's LSST, Euclid, SKA, and Roman. The resulting population of quads will sample the galaxy population more uniformly and would provide robust constraints on galaxy properties. The succeeding paper will utilize the full 7D space of lensing observables: three relative angles, three distance ratios, and the Einstein radius. We expect the discriminating power of $\eta$ to increase with dimensionality. From the correlations and initial tests shown here, we are optimistic that our novel model-free analysis will result in robust estimations of the observed lens galaxy population properties.

\section*{Acknowledgements}

Thank you to my colleagues and close friends: Lindsey Gordon, Derek Channa Perera, and Sarah Taft. May the webs of life always keep us close. Particular thanks to Io Miller for the numerous late night lensing discussions and for the helpful life advice. Thank you PG1115 for always being there.

\section*{Data Availability}

Data generated from this article will be shared upon reasonable request to the corresponding author.



\bibliographystyle{mnras}
\bibliography{bib} 



\appendix

\section{The Fundamental Surface of Quads} \label{sec:fsq}

\begin{figure*}
    \centering
    \includegraphics[width=0.9\textwidth]{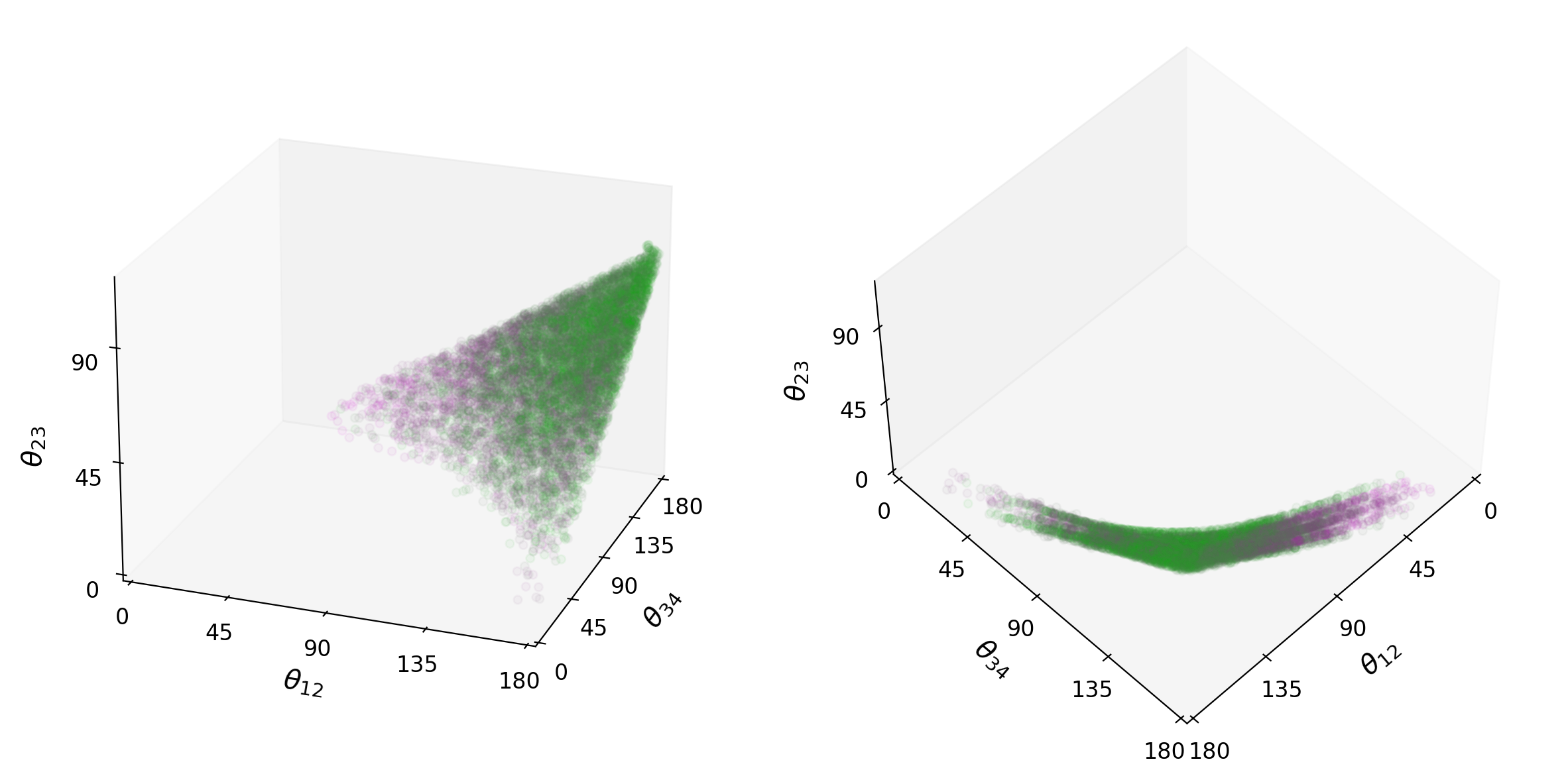}
    \caption{The 3-dimensional space of relative image angles, $\theta_{12}$, $\theta_{23}$, and $\theta_{34}$, with 10,000 quads from four ALPEIN-3 galaxies. The 2-dimensional surface produced by our quads resembles the FSQ. Each point corresponds to one quad. The points are color-coded based on $d_4/d_1$: the smaller $d_4/d_1$ values are magenta and larger $d_4/d_1$ values are green.}
    \label{fig:fsq}
\end{figure*}

The Fundamental Surface of Quads (FSQ) is a 2-dimensional surface created when plotting the three relative angles $\theta_{12}$, $\theta_{23}$, and $\theta_{34}$ from quads produced by the SIS+elliptical (or SISell) model \citep{2012woldesenbet}. The lensing potential of the SISell model is given by 
\begin{equation}
    \phi ~= rb[1+\gamma \cos(2\theta)],
\end{equation}
where $r$ and $\theta$ are polar coordinates in the lens plane, $b$ the normalization factor, and $\gamma$ the shear parameter. SISell models with different $b$ and $\gamma$ parameter values give rise to different points, or quads, on the FSQ and not different surfaces themselves. The 2D surface in the 3D space of relative image angles remains exactly invariant. When different elliptical lensing potentials and density profiles are used, the corresponding 2D surface deviates somewhat from the FSQ, but remains close to it. Therefore, quads generated by most elliptical density profiles, with arbitrary ellipticity and density profiles will lie very near the FSQ. This near-invariance makes FSQ a useful property in the analysis of quads.

The lower edge of the surface, when $\theta_{23}$ is small, corresponds to source positions approximately near the caustic. Conversely, the apex near ($\theta_{12}$, $\theta_{23}$, $\theta_{34}$) = (180\degree, 90\degree, 180\degree) corresponds to images in the Einstein cross configuration. 

The FSQ can be represented by $\theta_{23,\textup{FSQ}}$ as a function of $\theta_{12}$ and $\theta_{34}$, and is given by the unwieldy 4th order polynomial fit
\begin{multline} \label{eq:fsq}
    \theta_{23,\textup{FSQ}} = -5.792 + 1.783 \theta_{12} + 0.1648 \theta_{12}^2 \\ 
    - 0.04591 \theta_{12}^3  - 0.0001486 \theta_{12}^4  + 1.784\theta_{34} \\
    - 0.7275\theta_{34}\theta_{12} + 0.0549\theta_{34}\theta_{12}^2 + 0.01487\theta_{34}\theta_{12}^3 \\
    + 0.1643\theta_{34}^2 + 0.05493\theta_{34}^2\theta_{12} - 0.03429\theta_{34}^2\theta_{12}^2 \\
    - 0.04579\theta_{34}^3 + 0.01487\theta_{34}^3\theta_{12} - 0.0001593\theta_{34}^4,
\end{multline}
where the vertical proximity to the FSQ is given by
\begin{equation}
    \Delta\theta_{23} ~= ~\theta_{23} ~- ~\theta_{23,\textup{FSQ}}.
\end{equation}
The specific fit for $\theta_{23,\textup{FSQ}}$ in Eq. \ref{eq:fsq} is for the SIS+ell FSQ. That is, different elliptical mass models will result in slightly different polynomial fits to the FSQ. Therefore, systematic deviations from the FSQ can occur from quads generated from another smooth elliptical mass model other than the SISell model. However, these deviations were found to be small, typically $\Delta\theta_{23} < 0.1\degree$ or less.

The \textit{configuration invariant}, $\textup{I}_\textup{C}$, (see Eq. 2.6 of \citealt{1995kassiola}) is an invariant property of quadruply lensed quasars produced by simple lenses (i.e. lenses modelled by SISell). \citealt{1995kassiola} found that the angles between the direction of the semi-major axis of the lens, the lens center, and each of the four image positions (labelled $\phi_1$, $\phi_2$, $\phi_3$, and $\phi_4$, respectively) satisfy the \textit{configuration invariant}. That is, the quantity should vanish for any quad produced by a simple lens (i.e. $\textup{I}_\textup{C}$ = 0). Note that these angles are different from the FSQ angles defined in Section \ref{sec:lensobs} and Figure \ref{fig:pg1115}.

Recently, \citealt{2022falor} showed that the FSQ closely approximates the \textit{configuration invariant} when the invariant is described in the 3D space of lensing observables $\theta_{12}$, $\theta_{23}$, and $\theta_{34}$ (see their Figure 4). Therefore, it seems likely that the \textit{configuration invariant} explains the invariant nature of the FSQ. In other words, the FSQ is an "observer friendly" approximation of \textit{configuration invariant}. The difference between the two surfaces in the 3D space of lensing observables is negligible for the current paper and likely most use cases of the FSQ. 

Figure \ref{fig:fsq} shows the distribution of 10,000 quads from four ALPEIN-3 galaxies in the 3-dimensional space of relative image angles. The 2-dimensional surface produced by our quads resembles the FSQ, but with some $\theta_{23}$ thickness. The dispersion is caused by the four galaxies having non-elliptical asymmetry, and differing in their magnitude of asymmetry. A similar phenomena occurs with quads lens from an elliptical model with external shear \citep{2015woldesenbet}.

\section{Central Offset Tests} \label{sec:off}

\begin{figure*}
    \centering
    \includegraphics[width=0.8\textwidth]{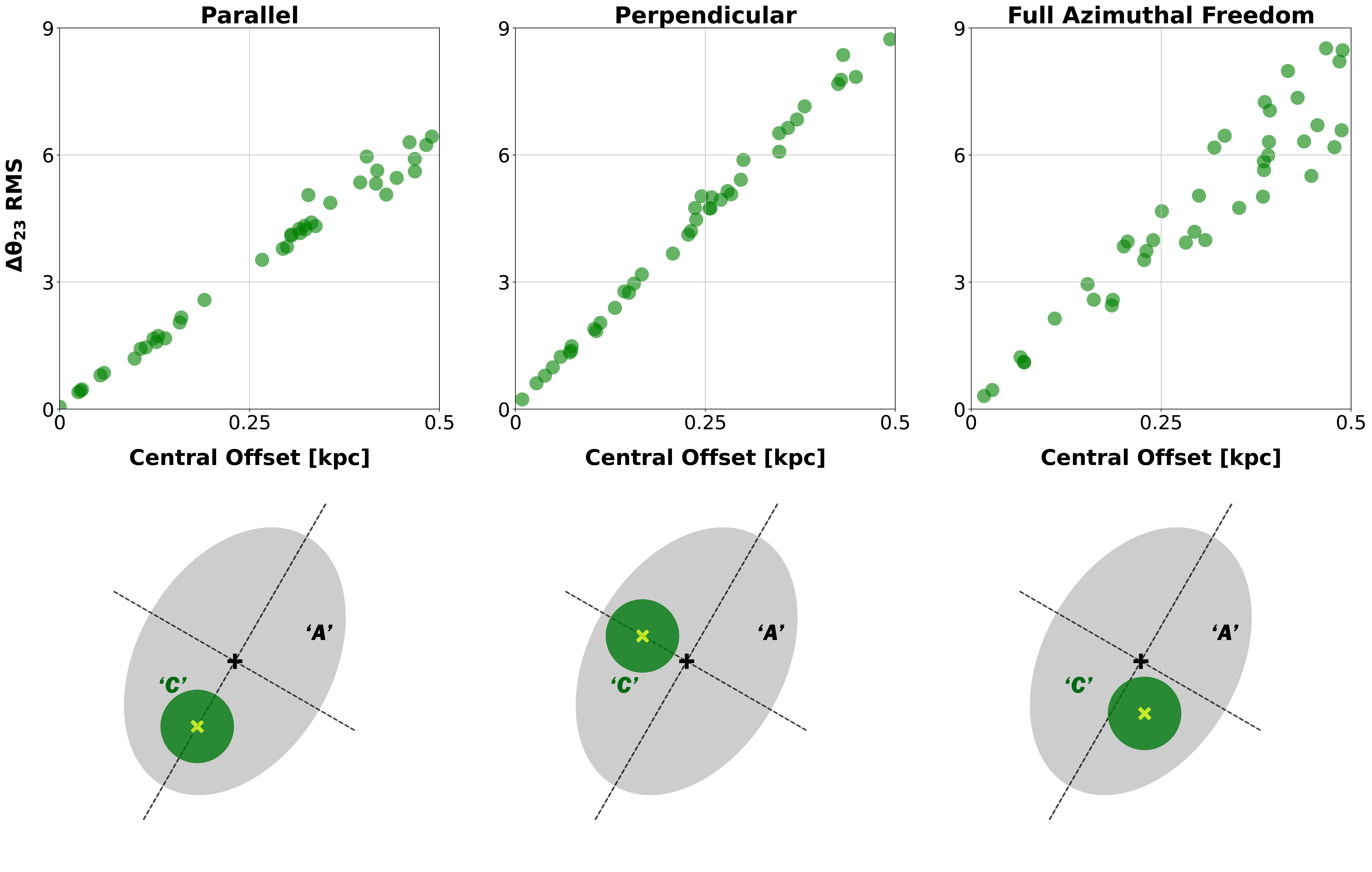}
    \caption{The RMS deviation of $\Delta\theta_{23}$ from zero plotted against the magnitude of the central offset between the dark matter component `A' and baryonic component `C' for three different orientations. The baryonic offset was allowed to vary parallel and perpendicular to the dark matter halo position angle, and with full azimuthal freedom. The diagrams on the bottom demonstrate the offset between the two matter components visually. Offsets are exaggerated for clarity.}
    \label{fig:offsets}
\end{figure*}

In Section \ref{sec:alp2} ALPEIN-2 galaxies were allowed to have non-zero central offsets between the main dark matter halo `A' and the baryonic component `C' to allow for lopsidedness in the mass distributions. The misalignment is determined by offsetting the baryonic component $0 - 0.5$ kpc radially and randomly in polar angle from the center of the dark matter component. To determine how this azimuthal asymmetry affects the lensing observables, specifically the deviation from the FSQ, $\Delta\theta_{23}$, three different types of galaxies were created: one where offsets vary only parallel to the main lens position angle (PA), another with offsets only perpendicular to the main lens PA, and one with full azimuthal freedom. 

The affect of a non-zero central offset on $\Delta\theta_{23}$ is quantified by its RMS deviation from zero. Figure \ref{fig:offsets} shows the RMS vs. central offset correlations for all three orientations, plus a visual representation of the offset orientation. The gray oval, labeled `A' with center `+', represents the dark matter component and the green circle, labeled `C' with center `x', represents the baryonic component. The dotted lines show the orientation and PA of the main dark matter component. All three orientations, parallel, perpendicular, and full azimuthal freedom, show strong linear correlations between $\Delta\theta_{23}$ and offset. Notably, we find that perpendicular offsets with the same radius as parallel offsets result in larger magnitudes of $\Delta\theta_{23}$. That is, perpendicular offsets create more asymmetries in mass distributions than parallel offsets, which is likely caused by the ellipticity of the dark matter component. The results for azimuthal freedom yield $\Delta\theta_{23}$ RMS values between the two correlations.


\bsp	
\label{lastpage}
\end{document}